\documentclass[11pt]{article}

\usepackage{graphicx}
\usepackage{newtxtext}
\usepackage{newtxmath}
\usepackage{natbib}
\usepackage{hyperref}
\hypersetup{
    colorlinks = true,
    urlcolor   = blue,
    citecolor  = black,
}

\newcommand{\RomanNumeralCaps}[1]
\linenumbers

\newcommand\pr{\ensuremath{\partial}}
\newcommand\Real{\mbox{Re}}
\newcommand\Imag{\mbox{Im}}
\newcommand\Rey{\mbox{\textit{Re}}}  % Reynolds number

\begin{document}

\title{The effect of boundary conditions on the stability of two-dimensional flows in an annulus with permeable boundary}

\author{Konstantin Ilin\footnote{Department of Mathematics, University of York,
Heslington, York YO10 5DD, UK. Email address for correspondence: konstantin.ilin@york.ac.uk} 
 \and Andrey Morgulis\footnote{Department of Mathematics, Mechanics and Computer Science, The Southern Federal University, Rostov-on-Don, and South Mathematical Institute, Vladikavkaz Center of RAS, Vladikavkaz,
Russian Federation}}
 \maketitle

\begin{abstract}
We consider the stability of two-dimensional viscous flows in an annulus with permeable boundary.
In the basic flow, the velocity has nonzero azimuthal and radial components, and
the direction of the radial flow can be from the inner cylinder to the outer one or vice versa.
In most earlier studies, all components of the velocity were assumed to be given on the entire boundary
of the flow domain. Our aim is to study the effect of different boundary conditions on the stability
of such flows. We focus on the following boundary conditions: at the inflow part if the boundary (which may be either inner or outer cylinder)
all components of the velocity are known; at the outflow part of the boundary (the other cylinder), the normal stress
and either the tangential velocity or the tangential stress are prescribed.
Both types of boundary conditions are relevant to certain real flows: the first one - to porous cylinders,
the second - to flows, where the fluid leaves the flow domain to an ambient fluid which is at rest.
It turns out that both sets of boundary conditions make the corresponding steady flows more unstable (compared with earlier works
where all components of the velocity are prescribed on the entire boundary). In particular, it is demonstrated that even
the classical (purely azimuthal) Couette-Taylor flow becomes unstable to two-dimensional perturbations if
one of the cylinders is porous and the normal stress (rather than normal velocity) is prescribed on that cylinder.
\end{abstract}

%\begin{keywords}

%\end{keywords}

{\bf MSC Codes }  76D05, 76E07

\section{Introduction}
\label{sec:intro}

In this paper we study the stability of steady two-dimensional viscous flows in an annulus between
two permeable circular cylinders. In the basic flow, the velocity has nonzero azimuthal and radial components, and
the direction of the radial flow can be from the inner cylinder to the outer one
(the diverging flow) or from the outer cylinder to the inner one
(the converging flow).
The stability of viscous flows
of this type has been studied by many authors
\cite[see][]{Bahl, Min, Johnson, Kolyshkin, Kolesov, Serre, Martinand, Gallet2010, Fujita, Kerswell, Martinand2017,IM2013a, IM2013b, IM2017,IM2020}.
Most papers were motivated by applications to dynamic filtration devices
\cite[see, e.g.,][]{Wron'ski, Beadoin} and vortex flow reactors \cite[see][and references therein]{Giordano}.
It was also argued by \cite{Gallet2010} and \cite{Kerswell} that such flows may have some relevance
to astrophysical flows in accretion discs \cite[see also][]{Kersale}. Similar inviscid flows have been also used as a
model of a flow in the vaneless diffuser of a radial pump \cite[for references, see][]{Tsujimoto,Ljevar,Guadagni}.

In all these papers (except the ones on the flow in vaneless diffusors),
all components of the velocity vector are prescribed on the permeable boundary of the flow domain. In what follows,
these boundary conditions and the corresponding
boundary-value problem will be called the \emph{reference boundary condition} and the \emph{reference problem}.
It is widely accepted that these boundary conditions are appropriate for flows bounded by porous walls.
This approach ignores the problem of modelling the flow in the porous medium and effectively assumes that this flow is given. Its big advantage
is that one needs to study only the flow outside the porous medium. However,
this also means that other boundary conditions may be relevant for flows bounded by porous walls, and
it is known that in problems with permeable boundaries, the stability properties of a flow can be strongly affected
by a change in boundary conditions \cite[see][]{Chomaz}. It is therefore natural
to raise the question: what is the effect of different boundary conditions on
of the stability of steady two-dimensional flows in an annulus with permeable boundary? The aim of the present paper
is to answer this question for two sets of boundary conditions, both of which are different from the reference conditions.

We focus on the following boundary conditions. At the inflow part of the boundary (the flow inlet), which is either inner or outer cylinder,
we specify all components of the velocity. At the outflow part (the flow outlet), represented by the other cylinder,
the viscous normal stress in the free fluid is balanced by a given pressure in the porous wall and, in addition to that,
either the tangential stress or the tangential velocity is prescribed. Since the normal stress contains the pressure, these two sets of boundary conditions will be referred to
as the \emph{pressure-stress} and \emph{pressure-no-slip} conditions.
Note that the only difference between the pressure-no-slip conditions and the reference conditions is that
the condition for the normal velocity at the outlet is replaced with the condition for  normal stress, while in the pressure-stress conditions
the no-slip condition is also replaced with the condition for tangential stress.
We argue in section 3 that both sets of boundary condition are no less relevant to real fluid flows than the reference conditions:
the pressure-no-slip conditions - to flows between porous cylinders, the pressure-stress conditions - to flows in vaneless diffusers.
Another reason to consider these conditions is that they appear in
computational fluid dynamics as boundary conditions on artificial boundaries, used to obtain a finite computational domain for problems,
originally formulated in infinite domains \citep[see, e.g.,][and references there]{Gresho, Heywood}. In particular, both sets of conditions arise
in a weak formulation of the Navier-Stokes equations \cite[see][p. 61]{Gunzburger}. Studies of the stability of flows bounded by
artificial boundaries may shed some light of the upstream influence of boundary conditions imposed on such boundaries.

It turns out that, for both types of conditions at the outlet, the corresponding boundary-value problems
formally reduce to the same inviscid problem for the Euler equations in the limit of high radial Reynolds number
(based on the radial velocity at the inner cylinder and its radius).
In the inviscid problem, the boundary conditions at the inlet remain the same (as those in the viscous problem), while
only the pressure is prescribed at the outlet. This suggests that in both viscous problems, an inviscid instability, similar to
that studied earlier \cite[see][]{IM2013a,Kerswell}, is likely to occur for sufficiently high radial Reynolds number.

For both types of viscous boundary conditions,
we investigate the linear stability of the steady diverging and converging flows. Numerical calculations show that for high radial
Reynolds numbers, the stability properties of the viscous flows are well described by the inviscid theory, while for small and moderate values
of the radial
Reynolds number the stability properties for both types of the outlet boundary conditions may be very different
from what was found in \cite{IM2013b} for the reference problem.
In particular, in the problem with the pressure-no-slip conditions, it turns out
that both diverging and converging flows are unstable at arbitrarily small radial Reynolds numbers provided the azimuthal velocity
at the inlet is much higher than the radial velocity. In this case, it is possible to construct an asymptotic approximation
of the linear stability
problem, which agrees with numerical results. An interesting byproduct of this asymptotic approximation is that a particular case of the classical
Couette-Taylor flow (with purely azimuthal basic flow),
where one cylinder is impermeable (for the fluid) and rotating and the other one is permeable and stationary,
turns out to be unstable to two-dimensional perturbations provided the normal stress condition (instead of the normal velocity condition)
at the outer cylinder is imposed. This is strikingly different from the classical Couette-Taylor flow which is
stable to two-dimensional perturbations. Another unexpected result, valid for both types of boundary conditions, is that
there are flow regimes where the converging flows are unstable even if the azimuthal velocity at the inlet is zero.

The paper is organised as follows. In section 2, the inviscid problem is considered. The effects of viscosity are analysed in section 3.
Section 4 contains the discussion of the results.

%%%%%%%%%%%%%%%%%%%%%%%%%%%%%%%%%%%%%%%%%%%%%%%%%%%%%%%%%%%%%%%%%%%%%%%%%%%%%%%%%%%%%%%%%%%%%%%%%%%%%%%%%%%%%%%%%%%%%%%%%%%%%%%%%%%%%

\setcounter{equation}{0}
\renewcommand{\theequation}{2.\arabic{equation}}

\section{Inviscid problem}\label{sec:inviscid}

\subsection{Formulation of the problem}\label{sec:problem}

 We consider two-dimensional inviscid incompressible flows in an annulus between two concentric circles
with radii $r_{1}$ and $r_{2}$ ($r_2 > r_1$). The circles are permeable for the fluid and there is a constant area flux $2\pi Q$ of
the fluid through the annulus. We shall call the flow \emph{diverging}
if the fluid is pumped into the annulus at the inner circle and taken out at the outer circle and \emph{converging} if the flow direction is reversed
(i.e. the fluid enters the annulus at the outer circle and leaves it at the inner one). Quantity $Q$ is positive for the diverging flow and negative
for the converging flow. For later use, we define the parameter
\[
\beta=\frac{Q}{\vert Q\vert},
\]
so that $ \beta=1$ for the converging flow and $ \beta=-1$ for the diverging flow.

Suppose that
$r_1$ is taken as a length scale, $r^2_{1}/\vert Q\vert$ as a time scale, $\vert Q\vert/r_{1}$ as a scale for the velocity and $\rho Q^2/r_{1}^2$ for the pressure
where $\rho$ is the fluid density. Then the two-dimensional Euler equations, written in non-dimensional variables, have the form
\begin{eqnarray}
&&u_{t}+ u u_{r} + \frac{v}{r}u_{\theta} -\frac{v^2}{r}= -p_{r} ,  \label{1} \\
&&v_{t}+ u v_{r} + \frac{v}{r}v_{\theta} +\frac{u v}{r}= -\frac{1}{r} \, p_{\theta} ,  \label{2} \\
&&\frac{1}{r}\left(r u\right)_{r} +\frac{1}{r} \, v_{\theta}=0.  \label{3}
\end{eqnarray}
Here $(r,\theta)$ are the polar coordinates, $u$ and $v$ are the radial and azimuthal components of the velocity and $p$ is the pressure.

If there is a non-zero flow of the fluid through the boundary, there are several sets of boundary conditions
on the parts of the boundary where the fluid enters the flow domain (the inlet) and leaves it (the outlet) which lead to
to mathematically correct initial-boundary-value problems \cite[for references, see][]{Monakh, MorgYud}.
One set of boundary conditions is where the normal and tangent components of the velocity is given at the inlet, but only
the normal component of the velocity is prescribed at the outlet.
It has been shown by Kazhikhov \cite[see Chapter 4 in][]{Monakh} that an initial-boundary-value problem
for the Euler equations with these boundary conditions
is a well-posed problem. In what follows we always consider the same boundary conditions at the inlet: both components
of the velocity are prescribed.
We shall refer to these conditions, supplemented with a condition for the normal component of velocity at the outlet,
as the \emph{normal velocity conditions}.

Here our focus is on a different set of boundary conditions:
at the flow inlet, we have the same conditions as before (both the normal and tangent components of the velocity are prescribed),
but at the outlet, the pressure is given instead of the normal velocity.
The Euler equations with these boundary conditions has been studied by \cite{Kazhikhov-Ragulin},
who have shown that the corresponding mathematical problem is well-posed.
We shall call these conditions the \emph{pressure conditions}.

Thus, our boundary conditions are
\begin{equation}
u\!\bigm\vert_{r=1}=1, \quad v\!\bigm\vert_{r=1}=\gamma_1,
\quad p\!\bigm\vert_{r=a}=p_0,  \label{4a}
\end{equation}
for the diverging flow ($\beta=1$) and
\begin{equation}
u\!\bigm\vert_{r=a}=-\frac{1}{a}, \quad v\!\bigm\vert_{r=a}=\frac{\gamma_2}{a},
\quad p\!\bigm\vert_{r=1}=p_0,  \label{4b}
\end{equation}
for the converging flow ($\beta=-1$). Here $a=r_2/r_1$, $p_0$ and $\gamma_{1,2}$ are constants ($p_0$ is the dimensionless pressure
at the outlet and $\gamma_{1,2}$ are the ratios of the azimuthal velocity to the radial velocity at the inner and outer cylinders, respectively)\footnote{In general, one can consider
non-constant $\gamma_{1,2}$ and $p_0$, i.e. given functions $\gamma_{1,2}(\theta,t)$ and $p_0(\theta,t)$, consistent with the restriction
that $u\vert_{r=a}>0$ if $\beta=1$ and $u\vert_{r=1}<0$ if $\beta=-1$ for all $\theta$ and $t$.}.

Equations (\ref{1})--(\ref{3}) with boundary conditions, given by either (\ref{4a}) or (\ref{4b}), have the following simple rotationally-symmetric solutions:
\begin{equation}
u(r,\theta)=\frac{\beta}{r}, \quad
v(r,\theta)=\left\{
\begin{array}{ll}
\gamma_1/r, &\hbox{if} \ \beta=1 \\
\gamma_2/r, &\hbox{if} \ \beta=-1
\end{array}\right. \label{5}
\end{equation}
with the pressure given by
\begin{equation}
p=p_0- \frac{1+\gamma_1^2}{2}\left(\frac{1}{r^2}-\frac{1}{a^2}\right)  \label{5a}
\end{equation}
for the diverging flow ($\beta=1$) and by
\begin{equation}
p=p_0+ \frac{1+\gamma_2^2}{2}\left(1-\frac{1}{r^2}\right)  \label{5b}
\end{equation}
for the converging flow ($\beta=-1$).
In the next section we investigate the stability of these steady flows.

\subsection{Inviscid stability analysis}\label{sec:inviscid_stability}

We consider a small perturbation
$(\tilde{u}, \tilde{v}, \tilde{p})$ in the form of the normal mode
\begin{equation}
\{\tilde{u}, \tilde{v}, \tilde{p}\} = \Real\left[\{\hat{u}(r), \hat{v}(r), \hat{p}(r)\} e^{\sigma t + in\theta}\right]  \label{3.1}
\end{equation}
where $n\in\mathbb{Z}$. This leads to the linearised equations:
\begin{eqnarray}
&&\left(\sigma +  \frac{in\gamma_{\alpha}}{r^2} + \frac{\beta}{r} \, \pr_{r} \right) \hat{u}
-\frac{\beta}{r^2} \, \hat{u} -\frac{2\gamma_{\alpha}}{r^2} \, \hat{v} = -\hat{p}_{r} ,  \label{3.2} \\
&&\left(\sigma +  \frac{in\gamma_{\alpha}}{r^2} + \frac{\beta}{r} \, \pr_{r} \right) \hat{v}
+\frac{\beta}{r^2} \, \hat{v}  = -\frac{in}{r} \, \hat{p} ,  \label{3.3} \\
&&\frac{1}{r}\left(r \hat{u}\right)_{r} +\frac{in}{r} \, \hat{v}=0, \label{3.4}
\end{eqnarray}
where $\gamma_{\alpha}=\gamma_1$ if $\beta=1$ and $\gamma_{\alpha}=\gamma_2$ if $\beta=-1$,
and the two sets of boundary conditions
\begin{equation}
\hat{u}(1)=0,
\quad \hat{v}(1)=0, \quad \hat{p}(a)=0 \label{linBCs_div}
\end{equation}
for the diverging flow ($\beta=1$) and
\begin{equation}
\hat{u}(a)=0,
\quad \hat{v}(a)=0, \quad \hat{p}(1)=0 \label{linBCs_conv}
\end{equation}
for the converging flow ($\beta=-1$).
Equations (\ref{3.2})--(\ref{3.4}) with either set of boundary conditions represent an eigenvalue problem for
$\sigma$.

First we note that both eigenvalue problems have no nontrivial solution for $n=0$. Indeed,
Eq. (\ref{3.4}) for $n=0$ and the boundary conditions for $\hat{u}$ imply that $\hat{u}=0$.
Equation (\ref{3.3}) yields $\hat{v} = C \, r^{-1} \, e^{-\sigma r^2/2\beta}$
where $C$ is an arbitrary constant. Substitution of this into the boundary condition
$\hat{v}(1)=0$ for $\beta=1$ or $\hat{v}(a)=0$ for $\beta=-1$ leads to the conclusion that $C=0$.
So, from now on we focus on eigenvalue problems with $n\neq 0$.

It is
convenient to introduce the stream function $\hat{\psi}(r)$ such that
\[
\hat{u}=\frac{in}{r} \, \hat{\psi}(r), \quad \hat{v}=-\hat{\psi}'(r).
\]
Eliminating the pressure from Eqs. (\ref{3.2}) and (\ref{3.3}), we obtain
\begin{equation}
\left(\sigma +  \frac{in\gamma_{\alpha}}{r^2} + \frac{\beta}{r} \, \pr_{r} \right)L \hat{\psi}=0 ,  \label{3.7} \\
\end{equation}
where
\begin{equation}
L \hat{\psi}= \hat{\psi}''+\frac{1}{r}\hat{\psi}'-\frac{n^2}{r^2}\hat{\psi} . \label{3.8}
\end{equation}
It follows from (\ref{3.7}) that
\begin{equation}
L \hat{\psi}= C e^{-\beta g_{\alpha}(r)} \label{3.9}
\end{equation}
where $C$ is a constant and $g_{\alpha}(r)=\sigma r^2/2+i n \gamma_{\alpha} \ln r$ for $\alpha=1,2$.
The general solution of (\ref{3.9}) can be written as
\begin{equation}
\hat{\psi} = \frac{C_{1}}{r^n}+C_{2} r^n + \frac{C}{2n}\int\limits_{1}^{r}
\left(r^n s^{-n+1}- r^{-n}s^{n+1}\right)e^{-\beta g_{\alpha}(s)} \,  ds \label{3.10}
\end{equation}
where $C$, $C_{1}$ and $C_{2}$ are arbitrary constants.

\subsubsection{Diverging flow ($\beta=1$)}\label{sec:inviscid_stability_div}

Now consider the diverging flow ($\beta=1$). In terms of $\hat{\psi}(r)$, the first two boundary conditions (\ref{linBCs_div}) take the form
\[
\hat{\psi}(1) = 0, \quad \hat{\psi}'(1) = 0 .
\]
Substitution of the general solution, given by (\ref{3.10}), into these conditions yields
$C_{1}=C_{2}=0$, so that Eq. (\ref{3.10}) simplifies to
\begin{equation}
\hat{\psi} = \frac{C}{2n}\int\limits_{1}^{r}
\left(r^n s^{-n+1}- r^{-n}s^{n+1}\right)e^{-g_1(s)} \,  ds \label{3.12}
\end{equation}
To satisfy the last of the boundary conditions (\ref{linBCs_div}), we employ Eq. (\ref{3.3}). As a result, we obtain
\begin{equation}
\left.\left(\sigma +  \frac{in\gamma_1}{r^2} + \frac{1}{r} \, \pr_{r} \right) \left(r \, \pr_{r}\hat{\psi}\right)\right\vert_{r=a}=0.  \label{3.13}
\end{equation}
Substitution of (\ref{3.12}) into (\ref{3.13}) yields the dispersion relation
for $\sigma$:
\begin{equation}
D(\sigma, n, a, \gamma_1)\equiv \left(a^2\sigma +in\gamma_1 +n\right) a^{n-3} \, I_1 + \left(a^2\sigma +in\gamma_1 -n\right) a^{-(n+3)} \, I_2
+ \frac{2}{a} \, e^{-g_1(a)}=0 \label{DR_div}
\end{equation}
where
\begin{equation}
I_1 = \int\limits_{1}^{a} r^{-n+1} \, e^{-g_1(r)} \, dr ,  \quad
I_2 = \int\limits_{1}^{a} r^{n+1} \, e^{-g_1(r)} \, dr . \label{3.14}
\end{equation}
Evidently, the dispersion relation has the following properties
\[
\overline{D(\sigma, n, a, \gamma_1)} = D(\overline{\sigma}, -n, a, \gamma_1), \quad
D(\sigma, -n, a, -\gamma_1) = D(\sigma, n, a, \gamma_1).
\]
(Here `bar' denotes complex conjugation.) These properties imply that we need to consider only positive $n$ and $\gamma_1$.

Numerical evaluation of the dispersion relation shows that there are no eigenvalues with positive real parts if $\gamma_1=0$. If $\gamma_1$ increases from $0$, the
roots of Eq. (\ref{DR_div}) move on the complex plane and, at some critical value, $\gamma_1=\gamma_{1cr}$, one of the eigenvalues crosses the imaginary axis,
so that for $\gamma_1 >\gamma_{1cr}$ there is an eigenvalue with positive real part, and hence, the flow is unstable.

\begin{figure}
\begin{center}
\includegraphics*[height=9cm]{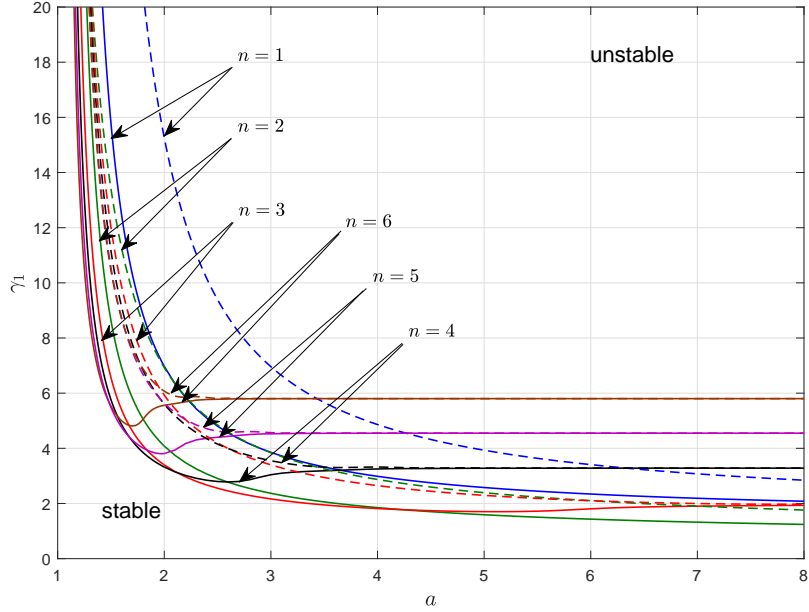} %Fig 1
\end{center}
%\vskip -10mm
\caption{Diverging flow: solid curves represent neutral curves for azimuthal modes with $n=1,\dots,6$; dashed curves show neutral curves for
the normal  velocity  condition, taken from \cite{IM2013a}.}
\label{stab_bound}
\end{figure}

Numerical calculations produced the stability diagram presented in Fig. \ref{stab_bound}.
It shows neutral curves on the $(a,\gamma_1)$ plane for normal modes with $n=1,\dots,6$. The instability region for each mode
is above the corresponding curve. Solid curves represent neutral curves for the pressure conditions.
Dashed curves show the results of \cite{IM2013a} for the normal velocity condition.
For all curves in Fig. \ref{stab_bound}, $\Imag(\sigma)\neq 0$. This means that the instability is oscillatory, and neutral modes
are periodic azimuthal waves.

Although the neutral curves in both problems
look qualitatively similar, there are two interesting differences, namely:
\begin{itemize}
\item[(i)] For each azimuthal mode, the curve for the problem considered in the present paper
is below the one corresponding to the normal velocity condition, which means that the same flow is more unstable if the pressure condition is used, and the gap between each pair of curves corresponding to the same $n$ is larger for smaller $a$.
\item[(ii)] For the problem
with the normal velocity condition, the critical value of $\gamma_1$ is a monotone decreasing function of $a$, for all azimuthal modes. However,
in the case of the pressure condition,  the neutral curves for modes with higher azimuthal wave numbers (for $n=3, \dots ,6$) have a local minimum,
and the minimum is attained at smaller values of $a$ for  higher $n$.

\end{itemize}

We should mention here a recent paper by \cite{Guadagni}. Motivated by an application to a flow in a radial vaneless diffusor,
the authors studied the stability of the diverging flow given by Eqs. (\ref{5}) and (\ref{5a}).
Although the paper contains several typos/errors (most notably, in the dispersion relation),
the neutral curves presented there seem to agree with the curves in Fig. \ref{stab_bound}.

\vskip 1mm
\noindent
\emph{Remark 1 (on the limit of weak radial flow).}
It can be shown that, in the limit $\gamma_1\to \infty$,
\[
\sigma=-in\gamma_1 + \gamma_1^{1/2}\left(\lambda + O(\gamma_1^{-1/2})\right).
\]
where $\lambda$ is a root of the equation
\[
\int\limits_{0}^{\infty}e^{-\lambda x+in x^2} \, dx = 0.
\]
The corresponding eigenfunction is given by
\[
\hat{\phi}(r)=\left[F(\xi)-\frac{F(0)r^n}{1+a^{2n}}\left(1+a^{2n}/r^{2n}\right)\right]
\]
where
\[
F(\xi)=\int\limits_{\xi}^{\infty}(x-\xi)e^{-\lambda x+in x^2} \, dx, \quad \xi=\gamma_1^{1/2}(r-1).
\]
The derivation of this approximation
simply repeats the arguments laid down in \cite{IM2013a} \cite[see also][]{Kerswell} for the case of the normal velocity
conditions.

It turns out that the leading-order approximations to $\sigma$ are the same for both types of boundary conditions.
This suggests that,
for $\gamma_1\gg 1$, the instability in both problems has the same mechanism. This, however, does not mean that
the change in the boundary condition at the outlet has little effect for all values of $\gamma_1$. Indeed, the difference in
the stability properties of the same flow in these two problems is considerable, as one can see in Fig. \ref{stab_bound}.

\subsubsection{Converging flow ($\beta=-1$)}\label{sec:inviscid_stability_conv}

For the converging flow, similar calculations yield the following dispersion relation:
\begin{equation}
\tilde{D}(\sigma, n, a, \gamma_2)\equiv \left(\sigma +in\gamma_2 -n\right) \, \tilde{I}_1 + \left(\sigma +in\gamma_2 +n\right) \, \tilde{I}_2
+ 2 \, e^{g_2(1)}=0 \label{DR_conv}
\end{equation}
where
\begin{equation}
\tilde{I}_1 = \int\limits_{1}^{a} r^{-n+1} \, e^{g_2(r)} \, dr ,  \quad
\tilde{I}_2 = \int\limits_{1}^{a} r^{n+1} \, e^{g_2(r)} \, dr . \label{3.23}
\end{equation}
Evidently, the dispersion relation has the same properties as those for the diverging flow:
\[
\overline{\tilde{D}(\sigma, n, a, \gamma_2)} = \tilde{D}(\overline{\sigma}, -n, a, \gamma_2), \quad
\tilde{D}(\sigma, -n, a, -\gamma_2) = \tilde{D}(\sigma, n, a, \gamma_2).
\]
Again, these imply that we need to consider only positive $n$ and $\gamma_2$.

\begin{figure}
\begin{center}
\includegraphics*[height=9cm]{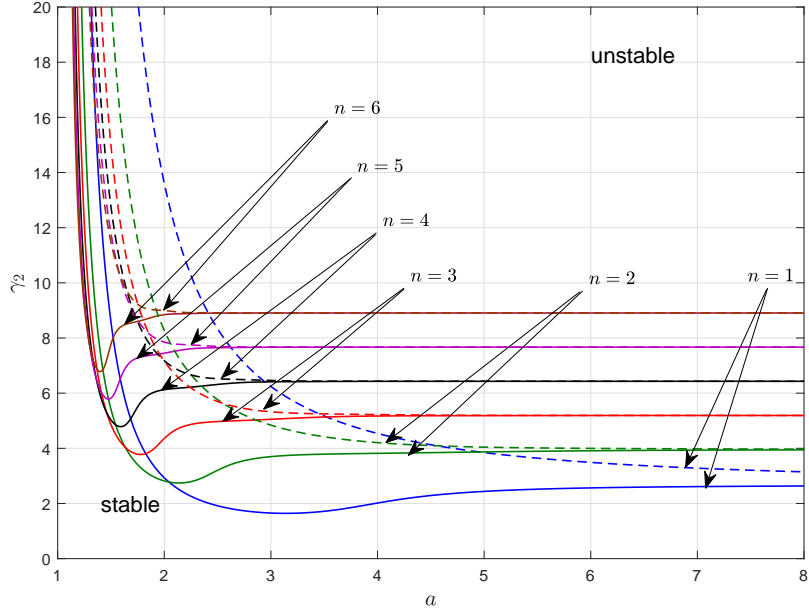} %Fig 2
\end{center}
%\vskip -10mm
\caption{Converging flow: solid curves represent neutral curves for azimuthal modes with $n=1,\dots,6$;
dashed curves show the results for the normal velocity  conditions.}
\label{stab_bound2}
\end{figure}

The neutral curves on the $(a,\gamma_2)$ plane for modes with $n=1,\dots,6$ are shown in Fig. \ref{stab_bound2}.
The instability region for each mode is above the corresponding curve. Again, the solid curves represent neutral curves for
the pressure conditions, and the dashed curves are curves for the normal velocity conditions computed in
\cite{IM2013a}. Qualitatively, the only difference between Figs. 1 and 2  is that every neutral curve in the latter has a local minimum.
Conclusions (i) and (ii) and the remark on the limit of weak radial flow made for the diverging flows are also true for the converging flows.

In the next section, we shall discuss the effects of viscosity.

%%%%%%%%%%%%%%%%%%%%%%%%%%%%%%%%%%%%%%%%%%%%%%%%%%%%%%%%%%%%%%%%%%%%%%%%%%%%%%%%%%%%%%%%%%%%%%%%%%%%%%%%%%%%%%%%%%%%%%%%%%
\setcounter{equation}{0}
\renewcommand{\theequation}{3.\arabic{equation}}

\section{Effects of viscosity}\label{sec:viscous_stability}

Here we consider two-dimensional viscous flows in an annulus with the pressure-stress and pressure-no-slip conditions.
The two-dimensional Navier-Stokes equations,
written using the same non-dimensional variables as those in section \ref{sec:problem}, have the form
\begin{eqnarray}
&&u_{t}+ u u_{r} + \frac{v}{r}u_{\theta} -\frac{v^2}{r}= -p_{r} +
\frac{1}{R} \left(\nabla^2 u-\frac{u}{r^2}-\frac{2}{r^2}v_{\theta}\right) ,  \label{4.1} \\
&&v_{t}+ u v_{r} + \frac{v}{r}v_{\theta} +\frac{u v}{r}= -\frac{1}{r} \, p_{\theta} +
\frac{1}{R} \left(\nabla^2 v-\frac{v}{r^2}+\frac{2}{r^2}u_{\theta}\right)  ,  \label{4.2} \\
&&\frac{1}{r}\left(r u\right)_{r} +\frac{1}{r} \, v_{\theta}=0,  \label{4.3}
\end{eqnarray}
where $R = \vert Q\vert/\nu$ is the radial Reynolds number  ($\nu$ is the kinematic viscosity of the fluid)
and $\nabla^2$ is the polar form of the Laplace operator:
\[
\nabla^2=\pr_r^2 + \frac{1}{r}\pr_r + \frac{1}{r^2}\pr_\theta^2 .
\]
Both components of the velocity are prescribed at the inlet:
\begin{equation}
u\!\bigm\vert_{r=1}=1, \quad v\!\bigm\vert_{r=1}=\gamma_1
  \label{4.4}
\end{equation}
for the diverging flow ($\beta=1$) and
\begin{equation}
u\!\bigm\vert_{r=a}=-\frac{1}{a}, \quad v\!\bigm\vert_{r=a}=\frac{\gamma_2}{a} \label{4.5}
\end{equation}
for the converging flow ($\beta=-1$).

The boundary conditions at the outlet are as follows.

\begin{itemize}

\item[(i)] \emph{The pressure-stress conditions:} for the diverging flows, these are given by
\begin{eqnarray}
&&\left.\left(-p+\frac{2}{R} u_r\right) \right\vert_{r=a}=-p_0,   \label{4.6} \\
&&\left.\frac{1}{R}\left(\frac{1}{r} \, u_{\theta}+v_r - \frac{1}{r} \, v\right) \right\vert_{r=a}= s_0,   \label{4.7}
\end{eqnarray}
and, for the converging flows, by
\begin{eqnarray}
&&\left.\left(-p+\frac{2}{R} u_r\right) \right\vert_{r=1}=-p_0,   \label{4.8} \\
&&\left.\frac{1}{R}\left(\frac{1}{r} \, u_{\theta}+v_r - \frac{1}{r} \, v\right) \right\vert_{r=1}= s_0,   \label{4.9}
\end{eqnarray}
where $p_0$ and $s_0$ are constants.
For the converging flow, the sign of $s_0$ in Eq. (\ref{4.9}) is chosen so as to make this condition look similar to
condition (\ref{4.7}). This means that $(-s_0)$ (not $s_0$ as in Eq. (\ref{4.7})) is the external tangential force (per unit area).

\item[(ii)] \emph{The pressure-no-slip conditions} are
\begin{equation}
\left.\left(-p+\frac{2}{R} u_r\right) \right\vert_{r=a}=-p_0,   \quad v\bigm\vert_{r=a}=\frac{\gamma_{2}}{a} \label{4.10}
\end{equation}
for the diverging flow and
\begin{equation}
\left.\left(-p+\frac{2}{R} u_r\right) \right\vert_{r=1}=-p_0,   \quad v\bigm\vert_{r=1}=\gamma_{1} \label{4.11}
\end{equation}
for the converging flow. Here $\gamma_{1,2}$ are the same non-dimensional parameters as before.

\end{itemize}

In the limit $R\to\infty$ conditions (\ref{4.6})--(\ref{4.9}) and (\ref{4.10}), (\ref{4.11}) reduce
to the inviscid boundary conditions of section 2.
This limit is not uniform as there is a viscous boundary layer at the outlet (but not at the inlet).
This is similar
to the case of the reference boundary conditions, for
which it is known that in the limit of high Reynolds number the boundary layer is formed at the outflow part of the boundary
\cite[see, e.g.,][]{Temam,Yudovich2001,Ilin2008}.

\vskip 1mm
\noindent
\emph{Remark 2 (on relevance of the pressure-no-slip and pressure-stress conditions to real flows).}
As was mentioned in section 1, both the reference boundary conditions and the conditions considered in this paper
can be used to model flows bounded by porous walls if we assume
that the flow in the porous walls is known.
In a more realistic model, one needs to solve the Navier-Stokes equations in the free flow domain and match it with a solution for
a flow in the porous medium of the walls (e.g. described by Darcy's law), using appropriate boundary conditions.
There are numerous papers on boundary conditions on an interface between a free fluid and porous medium
\cite[see, e.g.,][]{Joseph,Saffman,Haber}. There seems to be a consensus that the normal velocity and
the normal stress must be continuous across the interface. As for the tangential velocity,
either the no-slip condition (with the tangential velocity in the porous medium being zero) or the Beavers-Joseph
condition \cite[see][]{Joseph, Saffman} are used. In what follows, we assume that the porous medium is anisotropic,
with its permeability in the tangential direction being much smaller than the permeability in the normal direction,
so that the tangential velocity in the walls is very small and can be ignored. As a result, we have the no-slip condition for
the free fluid velocity. There are still two more conditions on the interface (for the normal velocity and the normal stress).
If we do not want to consider the flow in the porous medium and assume that it is known, one of these conditions
should be discarded in order to obtain a solvable mathematical problem for the Navier-Stokes equations.
The most common approach \cite[see][]{Bahl, Min, Serre, Martinand, Gallet2010, Fujita, Kerswell, Martinand2017, IM2013b, IM2020}
is to assume that the normal velocity in the porous medium is known
and discard the condition for the normal stress. This results in the reference conditions. However,
one can drop the normal velocity condition instead. The result will be the pressure-no-slip conditions.
Note that sometimes it is desirable to keep the condition for normal stress because it is
physically preferable to assume that the pressure (rather than the normal velocity) in the porous medium is known
or simply because it is less restrictive.
Indeed, suppose that we investigate the stability of purely azimuthal
flow between rotating porous cylinders to perturbations which do not change the boundary data. Then, if we use the reference
boundary conditions, the perturbation velocity will have to be zero at both cylinders and we end up with the stability
problem for the classical Couette-Taylor flow between rotating impermeable cylinders, so that the fact that the cylinders
are porous does not make any difference. However, if we prescribe the normal stress instead of the normal velocity,
there will be perturbations with nonzero normal velocity at the porous cylinders, which is more reasonable from the
physical viewpoint.

The pressure-stress condition can also be applicable to real flows. Consider, for example, a situation where the fluid leaves 
the flow domain to an ambient fluid which is at rest. In this case, it is natural to assume that the normal force at the outlet
is a force due to a constant pressure in the ambient fluid and that the tangential force is zero. For example,
these conditions can be used to model flows in vaneless diffusors of radial pumps \cite[see, e.g.,][] {Guadagni}.

\subsection{Basic flow}

Steady rotationally-symmetric flows whose stability we want to examine are given by
\begin{equation}
u=\frac{\beta}{r}, \quad v= V(r)=A \, r^{\beta R +1} + \frac{B}{r}  \label{4.12}
\end{equation}
where constants $A$ and $B$ are different for different boundary conditions at the outlet.
For both sets of boundary conditions, the pressure is
given by
\[
p=P(r)=p_0-\frac{2R^{-1}}{a^2}- \frac{1}{2}\left(\frac{1}{r^2}-\frac{1}{a^2}\right)-
\int\limits_{r}^{a}\frac{V^2(s)}{s} \, ds
\]
for the diverging flow ($\beta=1$) and
\[
p=P(r)=p_0+2R^{-1} + \frac{1}{2}\left(1-\frac{1}{r^2}\right)+
\int\limits_{1}^{r}\frac{V^2(s)}{s} \, ds
\]
for the converging flow ($\beta=-1$)

\vskip 2mm
\noindent
\emph{The pressure-stress conditions}.
Constants $A$ and $B$ are given by the following formulae:
\begin{equation}
A=\frac{(s_0 + 2\gamma_1 a^{-2}R^{-1})a^{-R}}{1+2R^{-1}a^{-(2+R)}}, \quad
B=\frac{\gamma_1-s_0 a^{-R}}{1+2R^{-1}a^{-(2+R)}} \label{4.13}
\end{equation}
for the diverging flow ($\beta=1$) and
\begin{equation}
A=-\frac{s_0 + 2\gamma_2 R^{-1}}{1-2R^{-1}a^{2-R}}, \quad
B=\frac{\gamma_2 + s_0 a^{2-R}}{1-2R^{-1}a^{2-R}} \label{4.14}
\end{equation}
for the converging flow ($\beta=-1$).
The steady solution formula for the converging flow ($\beta=-1$) is not defined for $R=2$. In this case,
the solution is given by
\begin{equation}
u=-\frac{1}{r}, \quad v= V(r)=\widetilde{A} \, \frac{\ln r}{r} + \frac{\widetilde{B}}{r}  \label{4.15}
\end{equation}
where
\[
\widetilde{A}=\frac{2\gamma_2 +s_0}{1+2\ln a}, \quad \widetilde{B}=\frac{\gamma_2 -2s_0\ln a}{1+2\ln a}.
\]
The dependence of the steady flow (\ref{4.12})--(\ref{4.14}) on $R$ is non-trivial and, for $R \gg 1$, it has a boundary layer
either at the outer cylinder (for the diverging flow) or at the inner one (for the converging flow).

It can be shown that, for $R\gg 1$, the azimuthal velocity profile is well approximated by the following asymptotic formula:
\[
V(r)=\left\{
\begin{array}{ll}
\gamma_1/r + s_0 a \, e^{-\eta} + O(R^{-1}) &\hbox{for} \ \beta=1 \\
\gamma_2/r - s_0 \, e^{-\xi} + O(R^{-1}) &\hbox{for} \ \beta=-1 \\
\end{array}\right.
\]
where $\xi=R(r-1)$ and $\eta=R(1-r/a)$. Note that if $s_0=0$ (or if $s_0\lesssim R^{-1}$ as $R\to\infty$), the above asymptotic
formula is different:
\[
V(r)=\left\{
\begin{array}{ll}
\gamma_1/r + \frac{2\gamma_1}{a} \, e^{-\eta} \, R^{-1} + O(R^{-2}) &\hbox{for} \ \beta=1 \\
\gamma_2/r - 2\gamma_2 \, e^{-\xi} \, R^{-1} + O(R^{-2}) &\hbox{for} \ \beta=-1 \\
\end{array}\right.
\]
which means that we have a weaker boundary layer.

\begin{figure}
\begin{center}
\includegraphics*[height=10cm]{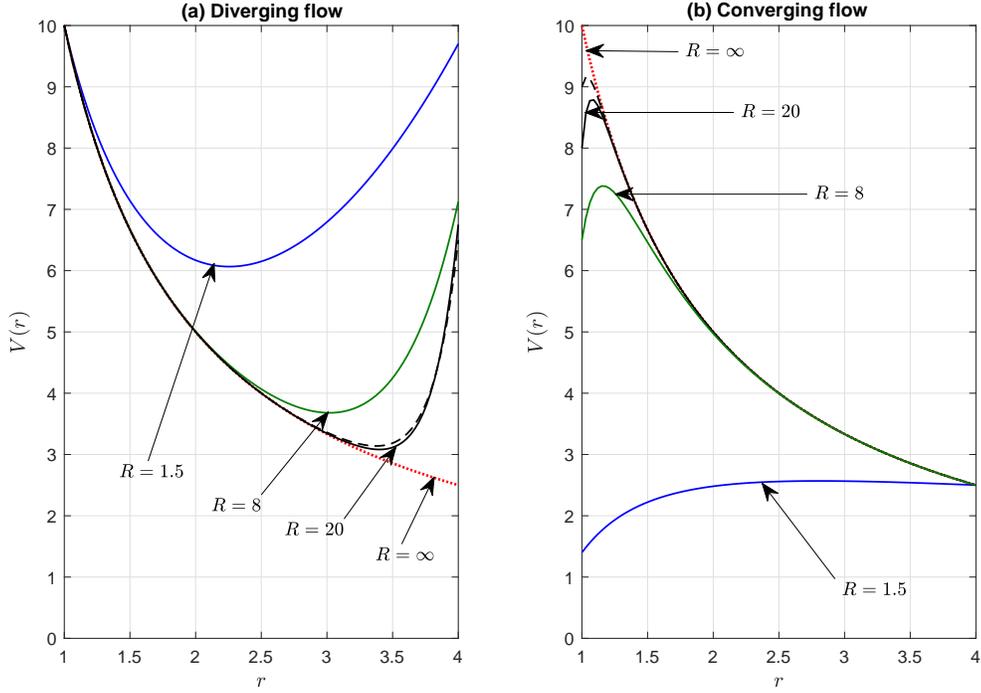} %Fig 3
\end{center}
%\vskip -10mm
\caption{Typical velocity profiles for $a=4$ and $R=1.5,8,20$.
(a) corresponds to the diverging flow ($\beta=1$) with $\gamma_1=10$ and $s_0=1$, (b) corresponds to
the converging flow ($\beta=-1$) with $\gamma_2=10$ and $s_0=1$. Dotted curves represent the inviscid velocity profiles.
Dashed curves show $V(r)$ computed using the asymptotic formulae ($R\gg 1$) for $R=20$.}
\label{Vel_pr_fig}
\end{figure}

Typical velocity profiles $V(r)$ for various $R$, as well as the corresponding asymptotic profiles given by the above formula,
are shown in Fig. \ref{Vel_pr_fig}. Evidently, the asymptotic formulae produce good approximations to the exact profile
even for $R=20$.

\vskip 1mm
\noindent
\emph{The pressure-no-slip conditions.}
In this case, the azimuthal velocity profile is the same as the one considered in \cite{IM2013b}.
Constants $A$ and $B$ can be written as
\begin{equation}
A=\frac{\gamma_2 -\gamma_1}{a^{\beta R+ 2}-1}, \quad B=\frac{a^{\beta R + 2}\gamma_1 -\gamma_2}{a^{\beta R + 2}-1}. \label{4.16}
\end{equation}
The steady solution depends on $\gamma_1$, $\gamma_2$ and $\beta R$ and is well defined for all $\beta R\neq -2$.
For $\beta R =-2$, the solution is given by Eq. (\ref{4.15}) with
\[
\widetilde{A}=\frac{\gamma_2 -\gamma_1}{\ln a}, \quad \widetilde{B}=\gamma_1.
\]
The asymptotic formula for $R\gg 1$ is
\[
V(r)=\left\{
\begin{array}{ll}
\gamma_1/r + ((\gamma_2-\gamma_1)/a) e^{-\eta} + O(R^{-1}) &\hbox{if} \ \beta=1 \\
\gamma_2/r - (\gamma_2-\gamma_1) e^{-\xi} + O(R^{-1}) &\hbox{if} \ \beta=-1 \\
\end{array}\right.
\]
where the boundary layer variables $\xi$ and $\eta$ are the same as before: $\xi=R(r-1)$ and $\eta=R(1-r/a)$.

From now on, we study the stability of the above steady flows.
The stability of steady flows in an annulus with
the reference boundary conditions have been studied in detail in \cite{IM2013b}. In what follows, all facts concerning the reference
boundary conditions are taken from that paper.

\subsection{Linear stability analysis}

Consider a small perturbation in the form (\ref{3.1}).
The linearised equations have the form
\begin{eqnarray}
&&\left(\sigma +  \frac{in V_{\beta}}{r} + \frac{\beta}{r} \, \pr_r \right) \hat{u}
-\frac{\beta}{r^2} \, \hat{u} -\frac{2V_{\beta}}{r} \, \hat{v} = - \pr_r \, \hat{p}  +
\frac{1}{R} \left(L \hat{u}-\frac{\hat{u}}{r^2}-\frac{2in}{r^2} \, \hat{v}\right) ,  \nonumber \\
&&\left(\sigma +  \frac{in V_{\beta}}{r} + \frac{\beta}{r} \, \pr_r \right) \hat{v}
+\frac{\beta}{r^2} \, \hat{v}  +\Omega_{\beta}(r) \hat{u} = -\frac{in}{r} \, \hat{p}  +
\frac{1}{R} \left(L \hat{v} -\frac{\hat{v}}{r^2}+\frac{2in}{r^2} \, \hat{u}\right),  \nonumber \\
&&\pr_r \left(r \hat{u}\right) +in \, \hat{v}=0, \label{4.17}
\end{eqnarray}
In Eqs. (\ref{4.17}), $V_{\beta}$ with $\beta=\pm 1$ is the azimuthal velocity for the diverging ($\beta=1$) and converging
($\beta=-1$) flows, and
\[
L  = \frac{d^2}{dr^2} + \frac{1}{r} \, \frac{d}{dr} - \frac{n^2}{r^2}, \quad \Omega_{\beta}(r)=V_{\beta}'(r)+\frac{V_{\beta}(r)}{r} .
\]
At the inlet, the boundary conditions for Eqs. (\ref{4.17}) are
\begin{equation}
\hat{u}(1)=0, \quad \hat{v}(1)=0 \label{4.18}
\end{equation}
for $\beta=1$ and
\begin{equation}
\hat{u}(a)=0, \quad \hat{v}(a)=0 \label{4.19}
\end{equation}
for $\beta=-1$.
At the outlet, the boundary conditions are either the pressure-stress conditions (that follow from Eqs. (\ref{4.6})--(\ref{4.9}))
\begin{equation}
\hat{p}(a)=\frac{2}{R} \, \hat{u}'(a), \quad \frac{in}{a} \, \hat{u}(a) + \hat{v}'(a) - \frac{1}{a} \, \hat{v}(a)=0 \label{4.20}
\end{equation}
for $\beta=1$ and
\begin{equation}
\hat{p}(1)=\frac{2}{R} \, \hat{u}'(1), \quad in \, \hat{u}(1) + \hat{v}'(1) -  \hat{v}(1)=0 \label{4.21}
\end{equation}
for $\beta=-1$, or the pressure-no-slip conditions
\begin{equation}
\hat{p}(a)=\frac{2}{R} \, \hat{u}'(a), \quad \hat{v}(a)=0 \label{4.22}
\end{equation}
for $\beta=1$ and
\begin{equation}
\hat{p}(1)=\frac{2}{R} \, \hat{u}'(1), \quad \hat{v}(1)=0 \label{4.23}
\end{equation}
for $\beta=-1$.

It  can be shown that if we restrict our analysis to axisymmetric perturbations,
then the basic steady flow (\ref{4.12}) is asymptotically stable
not only to small perturbations but also to perturbations of arbitrary amplitude. For the sake of completeness, the proof of
this fact is given in Appendix A. In particular, it implies that the mode with
$n=0$ cannot be unstable for any value of the Reynolds number. So, we shall consider only the modes with $n\neq 0$.

In terms of the steam function $\hat{\psi}(r)$, the first two equations (\ref{4.17}) are replaced by the vorticity equation
\begin{equation}
\left(\sigma +  \frac{in V_{\beta}}{r} + \frac{\beta}{r} \, \pr_r \right) L \hat{\psi}-
\frac{in}{r} \, \Omega_{\beta}'(r) \hat{\psi} = R^{-1}L^2 \hat{\psi}.   \label{4.24}
\end{equation}
The inlet boundary conditions become
\begin{eqnarray}
&&\hat{\psi}(1)=0, \quad \hat{\psi}'(1)=0 \quad \textrm{for} \ \ \beta=1, \label{4.25} \\
&&\hat{\psi}(a)=0, \quad \hat{\psi}'(a)=0 \quad \textrm{for} \ \ \beta=-1. \label{4.26}
\end{eqnarray}
To find the pressure, we employ the second equation (\ref{4.17}). As a result, we have
\[
\hat{p}= \frac{i r}{n R}\left(L \hat{\psi}_r -\frac{\hat{\psi}_r}{r^2}+\frac{2n^2}{r^3} \, \hat{\psi}\right)
-\frac{i r}{n}\left[\left(\sigma +  \frac{in V_{\beta}}{r} + \frac{\beta}{r} \, \pr_r \right) \hat{\psi}_r
+\frac{\beta}{r^2} \, \hat{\psi}_r  - \frac{in}{r} \, \Omega_{\beta}\, \hat{\psi}\right].
\]
So, the normal stress conditions at the outlet (the first equations in (\ref{4.20})--(\ref{4.23}))
can be written as
\begin{equation}
\left.\left[\frac{1}{R}\left(L \hat{\psi}_r -\frac{1+2n^2}{r^2}\, \hat{\psi}_r+\frac{4n^2}{r^3} \, \hat{\psi}\right)
-\left(\sigma + \frac{in V_{1}}{r} + \frac{1}{r} \, \pr_r \right) \hat{\psi}_r
-\frac{1}{r^2} \, \hat{\psi}_r  + \frac{in}{r} \, \Omega_{1}\, \hat{\psi}\right]\right\vert_{r=a} = 0
 \label{4.27}
 \end{equation}
for $\beta=1$ and
\begin{equation}
\left.\left[\frac{1}{R}\left(L \hat{\psi}_r -\frac{1+2n^2}{r^2}\, \hat{\psi}_r+\frac{4n^2}{r^3} \, \hat{\psi}\right)
-\left(\sigma + \frac{in V_{-1}}{r} - \frac{1}{r} \, \pr_r \right) \hat{\psi}_r
+\frac{1}{r^2} \, \hat{\psi}_r  + \frac{in}{r} \, \Omega_{-1}\, \hat{\psi}\right]\right\vert_{r=1} = 0
 \label{4.28}
 \end{equation}
for $\beta=-1$.
The tangent stress conditions at the outlet (the second equations in (\ref{4.20}) and (\ref{4.21})) take the form
\begin{eqnarray}
&&\hat{\psi}''(a)-\frac{1}{a}\, \hat{\psi}'(a) + \frac{n^2}{a^2}\, \hat{\psi}(a) =0 \quad \textrm{for} \ \ \beta=1, \label{4.29} \\
&&\hat{\psi}''(1)- \hat{\psi}'(1) + n^2 \hat{\psi}(1) =0 \quad \textrm{for} \ \ \beta=-1. \label{4.30}
\end{eqnarray}
The no-slip conditions at the outlet (given by the second equations in (\ref{4.22}) and (\ref{4.23})) become
\begin{equation}
\hat{\psi}'(a)=0  \quad \textrm{for} \ \ \beta=1  \quad \textrm{and} \quad
\hat{\psi}'(1)=0  \quad \textrm{for} \ \ \beta=-1 . \label{4.31}
\end{equation}
Note that, in view of (\ref{4.31}), conditions (\ref{4.27}) and (\ref{4.28})
simplify to
\begin{equation}
\left.\left[\frac{1}{R}\left(\hat{\psi}_{rrr} + \frac{1}{r}\, \hat{\psi}_{rr} +\frac{4n^2}{r^3} \, \hat{\psi}\right)
- \frac{1}{r} \, \hat{\psi}_{rr}  + \frac{in}{r} \, \Omega_{1}\, \hat{\psi}\right]\right\vert_{r=a} = 0  \label{4.32}
 \end{equation}
for $\beta=1$ and
\begin{equation}
\left.\left[\frac{1}{R}\left(\hat{\psi}_{rrr} + \frac{1}{r}\, \hat{\psi}_{rr} +\frac{4n^2}{r^3} \, \hat{\psi}\right)
+ \frac{1}{r} \, \hat{\psi}_{rr} + \frac{in}{r} \, \Omega_{-1}\, \hat{\psi}\right]\right\vert_{r=1} = 0
 \label{4.33}
 \end{equation}
for $\beta=-1$.

Simply by looking at Eq. (\ref{4.24})--(\ref{4.31}) and Eqs. (\ref{4.12})--(\ref{4.16}), one can deduce the following.
First, for a given $\beta$, an eigenvalue is a function of five parameters:
$\sigma=\sigma(a, n, \gamma_1, \gamma_2, R)$ in the case of the pressure-no-slip conditions;
$\sigma=\sigma(a, n, \gamma_{\alpha}, s_0, R)$, with $\alpha=1$ for $\beta=1$ and $\alpha=2$ for $\beta=-1$,
in the case of the pressure-stress conditions.  Second, if $\sigma(a, n, \gamma_1, \gamma_2, R)$ (or
$\sigma(a, n, \gamma_{\alpha}, s_0, R)$) is an eigenvalue, then so are
$\bar{\sigma}(a, -n, \gamma_1, \gamma_2, R)$
(or $\bar{\sigma}(a, -n, \gamma_{\alpha}, s_0, R)$) and
$\sigma(a, -n, -\gamma_1, -\gamma_2, R)$ (or $\sigma(a, -n, -\gamma_{\alpha}, -s_0, R)$).
Here $\bar{\sigma}$ is the complex conjugate of $\sigma$.
These properties imply that
it suffices to consider
only positive $n$ and, also, a certain symmetry of the neutral curves (which will be used later).

For $R\gg 1$, an asymptotic theory of the eigenvalue problems with the pressure-stress or pressure-no-slip conditions
can be developed along the same lines as in \cite{IM2013b}. In particular, it can be shown that
both problems reduce to the inviscid spectral problem of section \ref{sec:inviscid_stability}. This is a non-trivial property
because of the following two facts:
(i) the basic viscous flow depends on the Reynolds number $R$, and (ii) a single inviscid steady flow represents a vanishing viscosity limit for
continuous families of viscous steady flows (given by Eqs. (\ref{4.12})--(\ref{4.16})).

We shall not go into details of the asymptotic procedure here. Instead, we shall solve the viscous eigenvalue problems numerically.

%%%%%%%%%%%%%%%%%%%%%%%%%%%%%%%%%%%%%%%%%%%%%%%%%%%%%%%%%%%%%%%%%%%%%%%%%%%%%%%%%%%%%%%%%%

\subsection{Numerical results}

The eigenvalue problems with the pressure-stress and pressure-no-slip conditions are solved numerically using the Galerkin method
with polynomial basis functions based on Legendre polynomials. In the problem with the pressure-stress conditions,
the basis functions are chosen to satisfy the boundary conditions at the inlet,
given by Eqs. (\ref{4.25}) or (\ref{4.26}), and at the outlet by (\ref{4.29})
or (\ref{4.30}). So, $\hat{\psi}$ is approximated by
\[
\hat{\psi}(r)=\sum_{k=0}^{N}c_k \phi_k(x), \quad \quad x=-1+\frac{2}{a-1} \, (r-1),
\]
with basis functions $\phi_k(x)$, given by
\[
\phi_k(x)=\left\{
\begin{array}{ll}
(1+x)^2(x+\alpha_{k})P_k(x), &\hbox{for} \ \beta=1 \\
(1-x)^2(x+\alpha_{k})P_k(x), &\hbox{for} \ \beta=-1 \\
\end{array} \right.
\]
for $k=0,\dots,N$, where $P_k(x)$ is the Legendre polynomial of degree $k$ and $\alpha_{k}$ is a constant chosen
so as to satisfy (\ref{4.29}) or (\ref{4.30}) (note that conditions (\ref{4.25}) or (\ref{4.26}) are automatically satisfied).
The normal stress condition, given by Eqs. (\ref{4.27}) or (\ref{4.28}),
is satisfied using the $\tau$-method \cite[see, e.g.,][]{Gottlieb}.

In the case of the pressure-no-slip conditions, the $\tau$-method yields spurious eigenvalues because
conditions (\ref{4.32}) and (\ref{4.33}) do not contain the spectral parameter $\sigma$. However, the same fact makes it
possible to construct basis functions which
satisfy all the boundary conditions.
These basis functions have the form
\[
\phi_k(x)=\left\{
\begin{array}{ll}
(1+x)^2(x^2+\alpha_{k} x + \beta_k)P_k(x), &\hbox{for} \ \beta=1 \\
(1-x)^2(x^2+\alpha_{k} x + \beta_k)P_k(x), &\hbox{for} \ \beta=-1 \\
\end{array} \right.
\]
where constants $\alpha_{k}$ and $\beta_{k}$ are chosen to satisfy the conditions given by the first equation (\ref{4.31})
and Eq. (\ref{4.32}) if $\beta=1$
and the second equation (\ref{4.31}) and Eq. (\ref{4.33}) if $\beta=-1$.

To verify the method, some of the computed eigenvalues were compared with eigenvalues obtained using the
shooting method. A further verification was provided by checking the consistency
of the results for high radial Reynolds numbers with the inviscid theory of section 2.
\begin{figure}
\begin{center}
\includegraphics*[height=8cm]{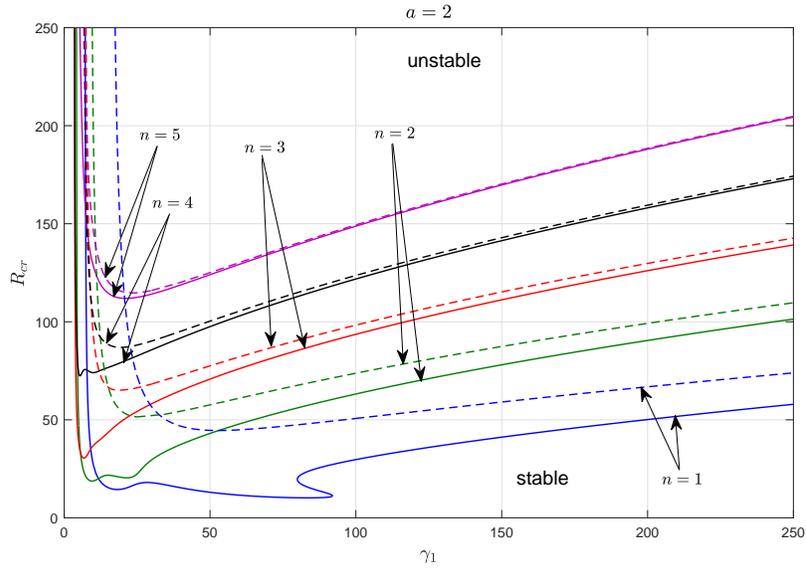} %Fig 4
\end{center}
%\vskip -10mm
\caption{Diverging flow with the pressure-stress conditions: critical $R$ versus $\gamma_1$ for $a=2$ and $n=1,\dots,5$.
Solid curves correspond to the pressure-stress conditions with $s_0=0$, dashed curves - to the reference conditions
with $\gamma_2=0$.}
\label{R_cr_vs_gamma_s0_a2_ALL_n}
\end{figure}

\begin{figure}
\begin{center}
\includegraphics*[height=8cm]{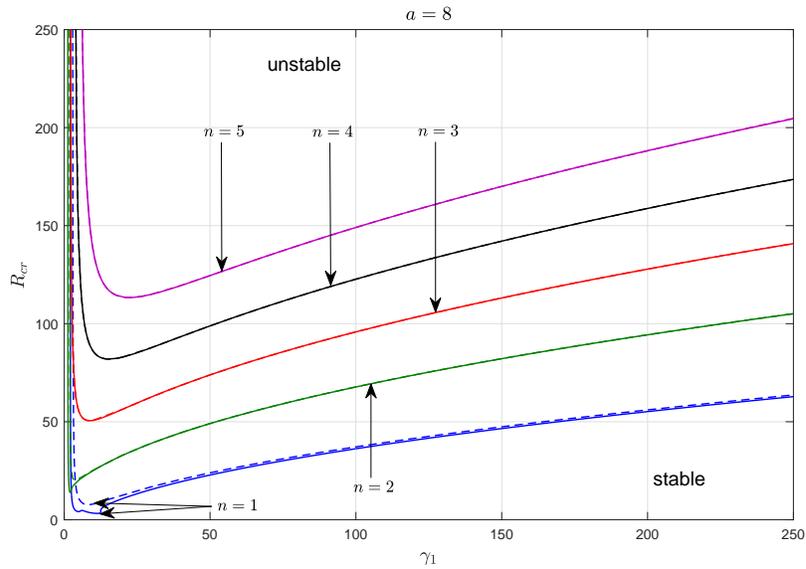}  %Fig 5
\end{center}
%\vskip -10mm
\caption{Diverging flow with the pressure-stress conditions: critical $R$ versus $\gamma_1$ for $a=8$ and $n=1,\dots,5$.
Solid curves correspond to the pressure-stress boundary conditions with $s_0=0$, dashed curves - for the reference conditions
with $\gamma_2=0$.}
\label{R_cr_vs_gamma_s0_a8_ALL_n}
\end{figure}

\subsubsection{Problem with the pressure-stress conditions.}

\emph{Diverging flow}.
Figures \ref{R_cr_vs_gamma_s0_a2_ALL_n} and \ref{R_cr_vs_gamma_s0_a8_ALL_n}
show neutral curves on the $(\gamma_1,R)$ plane for $a=2$ and $a=8$, respectively. The solid curves
represent critical values of $R$ as functions of $\gamma_1$ for modes with $n=1, \dots,5$ for the pressure-stress conditions
with $s_0=0$. The dashed curves are critical curves for the reference boundary conditions with $\gamma_2=0$, taken from \cite{IM2013b}.
All curves in Figs. \ref{R_cr_vs_gamma_s0_a2_ALL_n} and \ref{R_cr_vs_gamma_s0_a8_ALL_n} approach vertical asymptotes as
$\gamma_1$ tends to $\gamma_{1}^{*}(a,n)$ from the right, where $\gamma_{1}^{*}(a,n)$
is the critical value of $\gamma_1$ for the inviscid mode with azimuthal number $n$.
Numbers $\gamma_{1}^{*}(a,n)$ can be determined from the inviscid diagram shown in Fig. \ref{stab_bound}.

We note in passing that, in view of the symmetry properties of the eigenvalue problem,
critical curves for negative $\gamma_1$ can be obtained by reflecting the curves in
Figs. \ref{R_cr_vs_gamma_s0_a2_ALL_n} and \ref{R_cr_vs_gamma_s0_a8_ALL_n} about the vertical axis.

Figure \ref{R_cr_vs_gamma_s0_a2_ALL_n} shows that the critical curves for $a=2$ are below the corresponding curves for the problem with
the reference boundary conditions, and the gap between the curves with the same azimuthal wave number $n$ is larger for smaller
$n$ and decreases when $n$ increases. The same is true for the critical curves for $a=8$, but the effect is much weaker: one can see
in Fig. \ref{R_cr_vs_gamma_s0_a8_ALL_n} that the gap between the curves
with $n=1$ is much smaller then the corresponding gap for $a=2$, and it becomes invisible for modes with $n>1$. We can conclude that
the pressure-stress boundary conditions make the flow more unstable, and this effect is stronger for smaller $a$. The latter is not surprising, as
it is natural to expect that for wider annuli, the effect of the boundary conditions at the outlet is weaker.

\begin{figure}
\begin{center}
\includegraphics*[height=8cm]{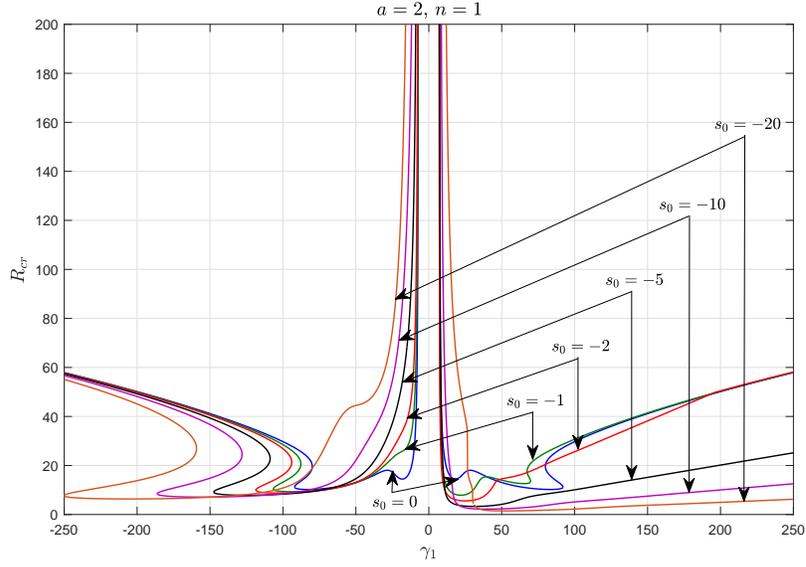} %Fig 6
\end{center}
%\vskip -10mm
\caption{Diverging flow with the pressure-stress conditions: critical $R$ versus $\gamma_1$ for $a=2$, $n=1$ and several values of $s_0$.}
\label{R_cr_a2_n1_ALL_s_fig}
\end{figure}

Figure \ref{R_cr_a2_n1_ALL_s_fig} shows critical curves for $a=2$, $n=1$ and several values of $s_0$.
While the curve for $s_0=0$ is symmetric
relative to the vertical axis, the curves for $s_0\neq 0$ are not symmetric. However, due to the symmetries of the eigenvalue problem
mentioned earlier, the critical curves for the same $a$ and $n$ and for $s_0=0$, $1$, $2$, $5$, $10$, $20$ can be obtained by reflecting the curves in
Fig. \ref{R_cr_a2_n1_ALL_s_fig} about the vertical axis $\gamma_1=0$.
Note that most neutral curves in Fig. \ref{R_cr_a2_n1_ALL_s_fig} have folds. This manifests itself especially clearly
in the left half of the figure (where the directions of the tangent velocity at the inlet and tangent force at the outlet coincide).
This means that for some fixed values of $\gamma_1$, as the radial Reynolds number increases, the flow becomes unstable, then stable, then unstable
again.

\begin{figure}
\begin{center}
\includegraphics*[height=8cm]{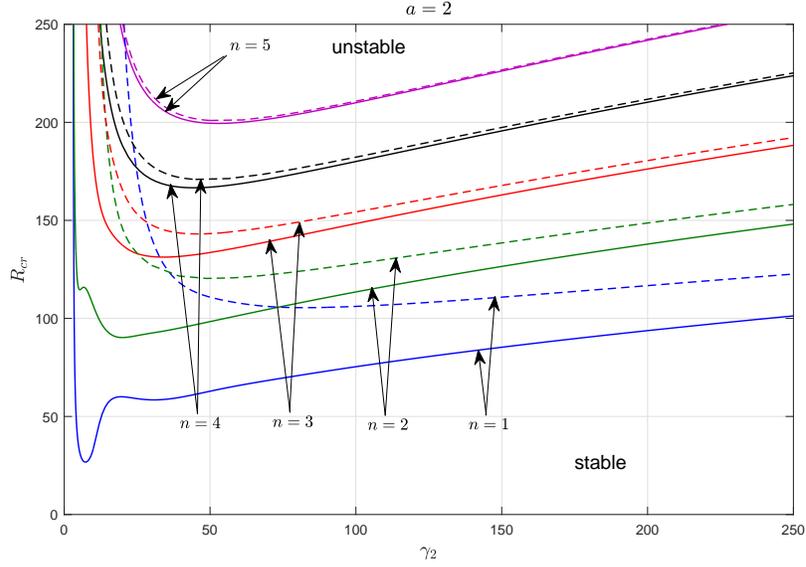} %Fig 7
\end{center}
%\vskip -10mm
\caption{Converging flow with the pressure-stress conditions: critical $R$ versus $\gamma_2$ for $a=2$ and $n=1,\dots,5$.
Solid curves correspond to the pressure-stress conditions with $s_0=0$, dashed curves - to the reference boundary conditions
with $\gamma_1=0$.}
\label{cr_Re_v_gam2_a2_ALL_n}
\end{figure}

\begin{figure}
\begin{center}
\includegraphics*[height=8cm]{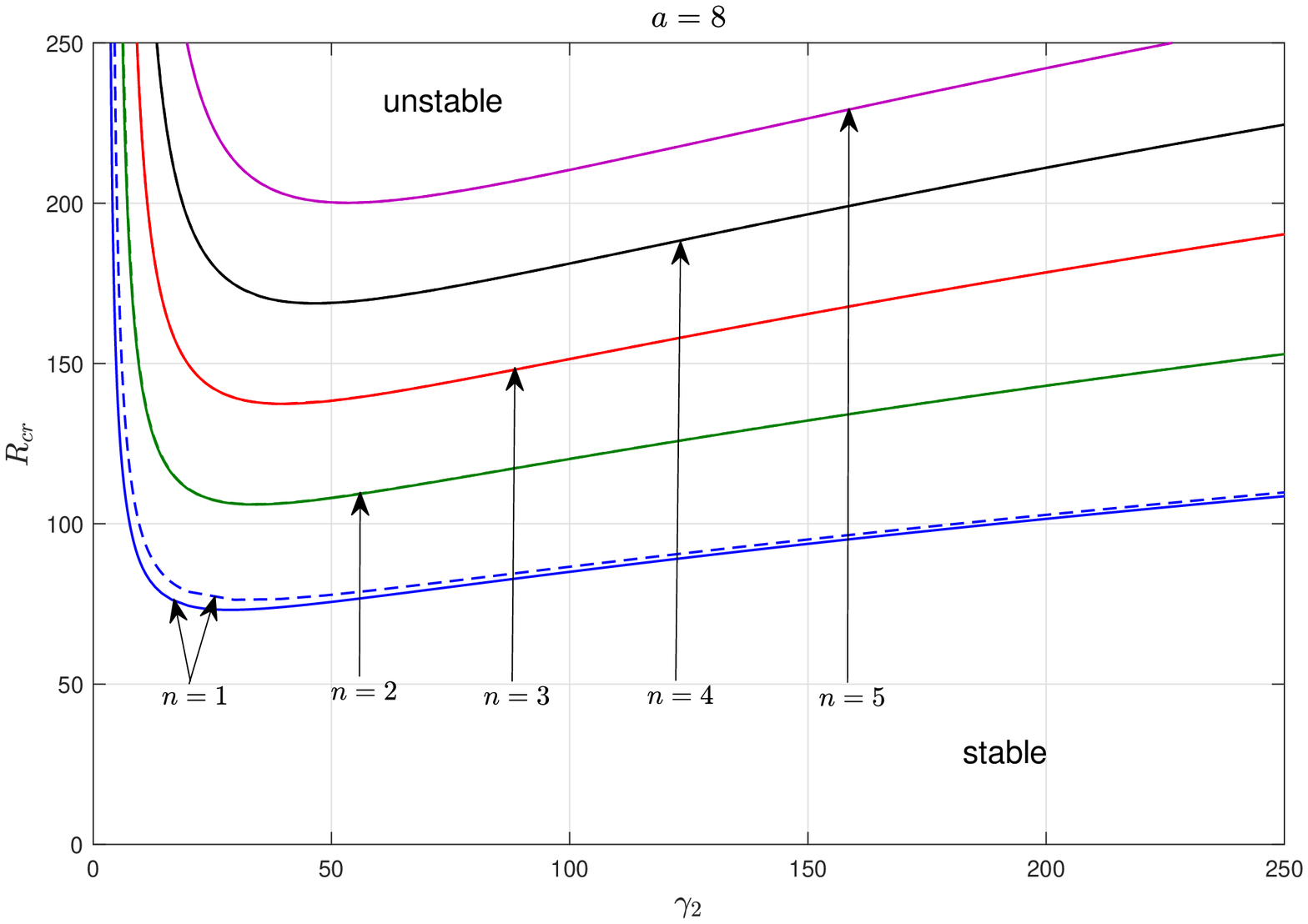}  %Fig 8
\end{center}
%\vskip -10mm
\caption{Converging flow with the pressure-stress conditions: critical $R$ versus $\gamma_2$ for $a=8$ and $n=1,\dots,5$.
Solid curves correspond to the pressure-stress boundary conditions with $s_0=0$; dashed curves - to the reference boundary conditions
with $\gamma_1=0$.}
\label{cr_Re_v_gam2_a8_ALL_n}
\end{figure}

\vskip 1mm
\noindent
\emph{Converging flow}.
Figures \ref{cr_Re_v_gam2_a2_ALL_n} and \ref{cr_Re_v_gam2_a8_ALL_n}
show critical curves on the $(\gamma_2,R)$ plane for the converging flow ($\beta=-1$) for $a=2$ and $a=8$.
The solid curves correspond to the pressure-stress conditions with $s_0=0$, and the dashed curves - to
the reference boundary conditions with $\gamma_1=0$.
Everything that has been said about the diverging flows can also be said
about the neutral curves for the converging flows. The only difference is that the critical curves for the converging flows
are above the corresponding curves for the diverging flows, i.e. the former are more stable than the latter, but we still have
the result that the flows with the pressure-stress conditions are more
unstable than those with the reference conditions.

\begin{figure}
\begin{center}
\includegraphics*[height=8cm]{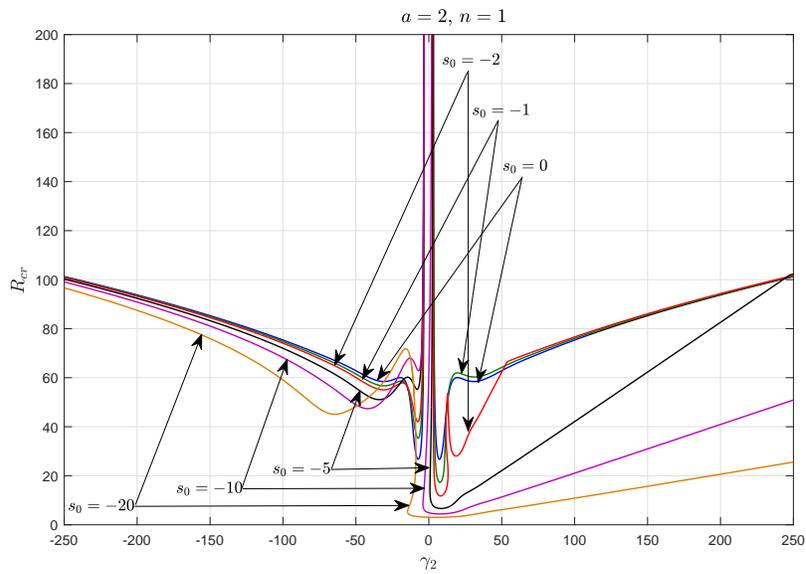}  %Fig 9
\end{center}
%\vskip -10mm
\caption{Converging flow with the pressure-stress conditions: critical $R$ versus $\gamma_2$ for $a=2$, $n=1$ and several values of $s_0$.}
\label{cr_R_a_2_n_1_ALL_s0_conv_fig}
\end{figure}

Figure \ref{cr_R_a_2_n_1_ALL_s0_conv_fig} shows critical curves for the converging flows for $a=2$, $n=1$ and several values of $s_0$.
The curve for $s_0=0$ is symmetric
relative to the vertical axis, but the symmetry is lost for nonzero $s_0$. Again,
the critical curves for the same $a$ and $n$ and for $s_0=0$, $1$, $2$, $5$, $10$, $20$ can be obtained by reflecting the curves in
Fig. \ref{cr_R_a_2_n_1_ALL_s0_conv_fig} about the vertical axis $\gamma_2=0$.
Note the oscillatory behaviour of the curves near the axis $\gamma_2=0$, which implies that, for certain fixed values of $R$, the stability properties
change a few times, when $\gamma_2$ increases.
For very high $R$ ($R\to\infty$), all critical curves approach the vertical asymptotes $\gamma_2=\pm\gamma_2^{*}(a,n)$
irrespective of values of $s_0$, where $\gamma_2^{*}(a,n)$ is the iniviscid instability boundary on the $(a,\gamma_2)$ plane,
shown in Fig. \ref{stab_bound2}.

There is an interesting feature in Fig. \ref{cr_R_a_2_n_1_ALL_s0_conv_fig} (which is absent in Fig. \ref{R_cr_a2_n1_ALL_s_fig}), namely:
the curves for $s_0=-10$ and $s_0=-20$ cross the vertical axis $\gamma_2=0$.
More precisely, for the converging flows with sufficiently large (in magnitude) $s_0$, there is a finite interval in $R$ where the flow
is unstable even for $\gamma_2=0$ (i.e. for purely radial flow at the inlet). For $s_0=\pm 10$, this instability interval is $R\in (4.907, 61.417)$; for $s_0=\pm 20$, it is $R\in (3.064 ,243.605)$.

%%%%%%%%%%%%%%%%%%%%%%%%%%%%%%%%%%%%%%%%%%%%%%%%%%%%%%%%%%%%%%%%%%%%%%%%%%%%%%%%

\begin{figure}
\begin{center}
\includegraphics*[height=8cm]{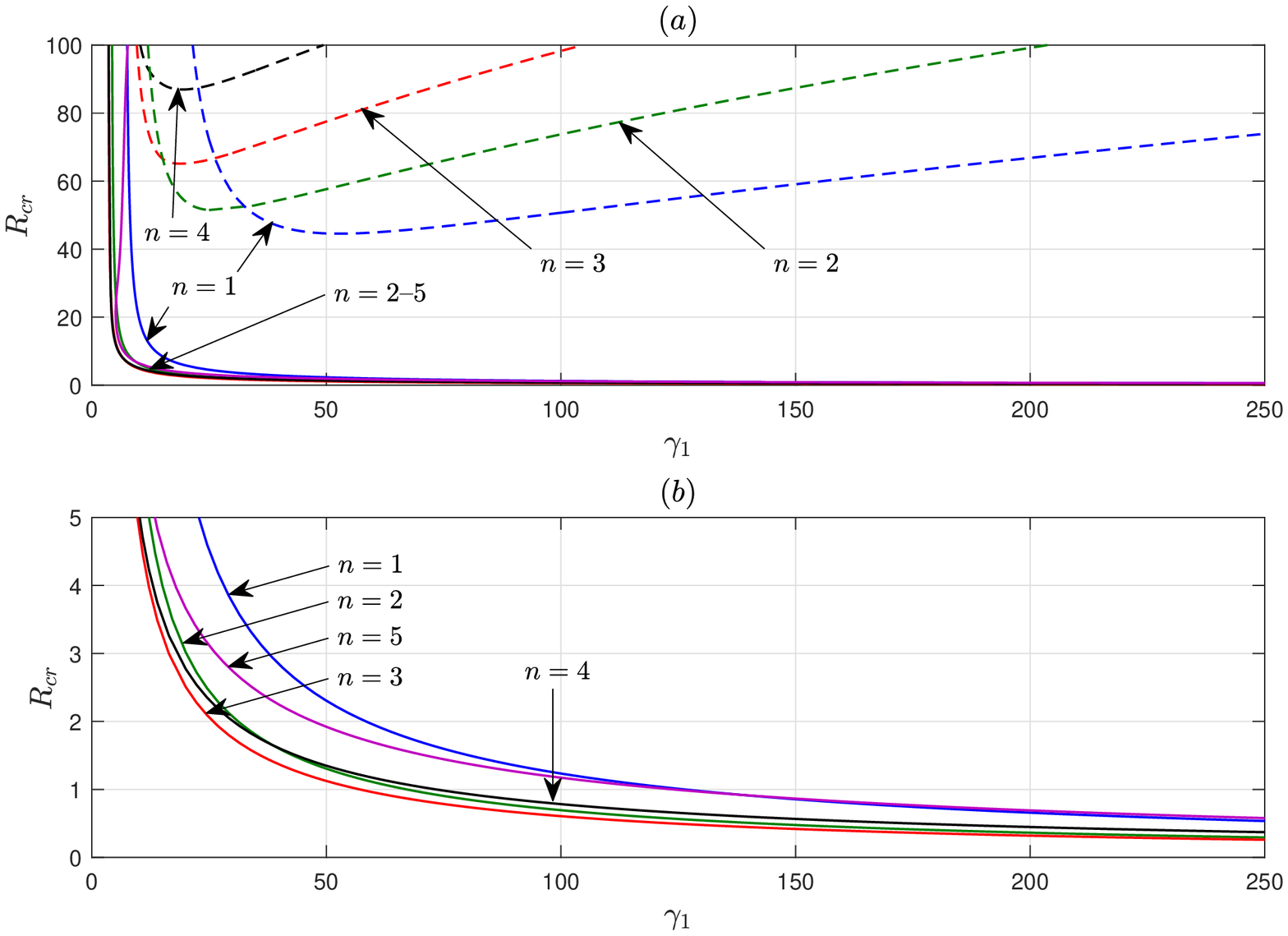}  %Fig 10
\end{center}
%\vskip -10mm
\caption{Diverging flow with the pressure-no-slip conditions: critical $R$ versus $\gamma_1$ for $a=2$, $\gamma_2=0$ and $n=1,\dots,5$;
(a) solid curves correspond to the pressure-no-slip conditions, dashed curves - to the reference boundary conditions;
(b) shows a magnified lower part of (a).}
\label{cr_Re_tang_vel_cond_a2_FIG}
\end{figure}

\begin{figure}
\begin{center}
\includegraphics*[height=8cm]{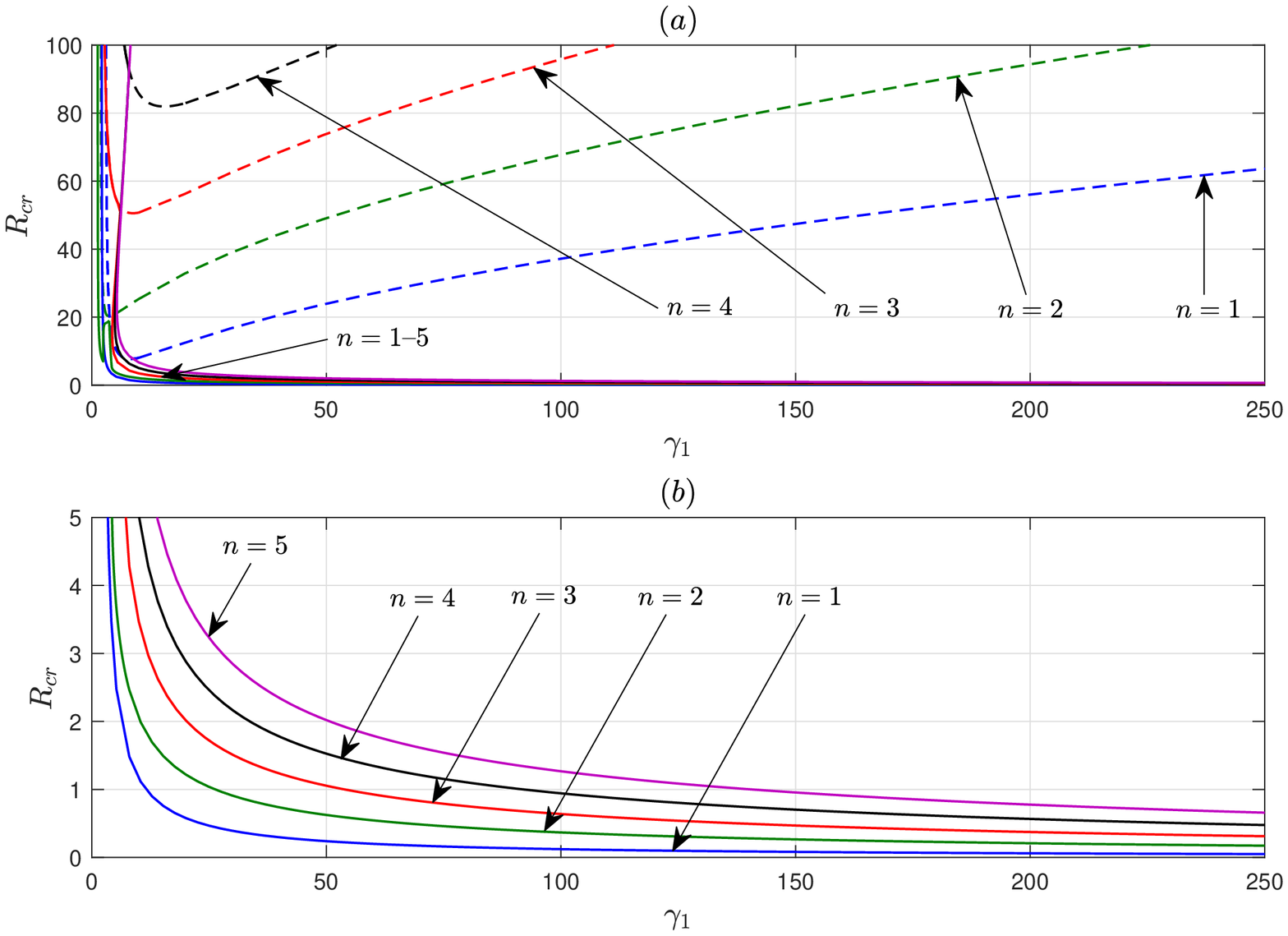}  %Fig 11
\end{center}
%\vskip -10mm
\caption{Diverging flow with the pressure-no-slip conditions: critical $R$ versus $\gamma_1$ for $a=8$, $\gamma_2=0$ and $n=1,\dots,5$;
(a) solid curves correspond to the pressure-no-slip conditions, dashed curves - to the reference boundary conditions;
(b) shows a magnified lower part of (a).}
\label{cr_Re_tang_vel_cond_a8_FIG}
\end{figure}

\subsubsection{Problem with the pressure-no-slip conditions.}

\emph{Diverging flow}.
The critical curves for the pressure-no-slip conditions (solid curves), as well as the curves for
the reference conditions (dashed curves), are shown in Figs. \ref{cr_Re_tang_vel_cond_a2_FIG}(a) for $a=2$ and \ref{cr_Re_tang_vel_cond_a8_FIG}(a) for $a=8$.
Figures \ref{cr_Re_tang_vel_cond_a2_FIG}(b) and \ref{cr_Re_tang_vel_cond_a8_FIG}(b) show the same curves
for small $R$ in more detail. Evidently, as $\gamma_1\to\infty$, the neutral curves monotonically approach the horizontal line $R=0$,
which suggests that
in the limit $\gamma_1\to \infty$ (equivalently, in the limit of a weak radial flow), the basic flow is unstable for all $R>0$.
This behaviour is very different from both the case of the pressure-stress conditions and the
case of the reference conditions, for which the critical Reynolds number grows linearly with $\gamma_1$ for $\gamma_1\gg 1$.

\begin{figure}
\begin{center}
\includegraphics*[height=8cm]{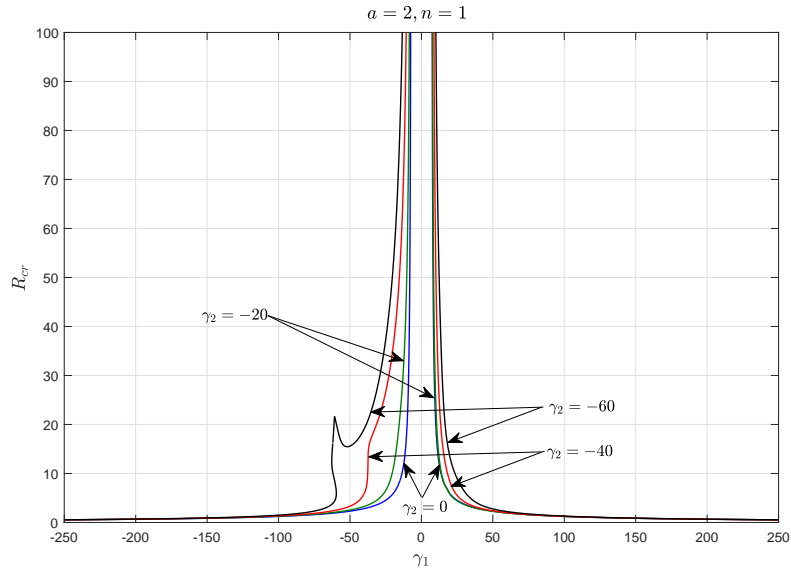}  %Fig 12
\end{center}
%\vskip -10mm
\caption{Diverging flow with the pressure-no-slip conditions: critical $R$ versus $\gamma_1$ for $a=2$, $n=1$ and $\gamma_2=0, -20, -40, -60$.}
\label{cr_Re_a2_n1_ALL_gam2_NEW_fig}
\end{figure}

The critical curves for $a=2$, $n=1$ and several values of $\gamma_2$ are shown in Fig. \ref{cr_Re_a2_n1_ALL_gam2_NEW_fig}.
Again, the critical curves for the same $a$ and $n$ and for $\gamma_2=0, 20, 40, 60$  can be obtained by reflecting the curves in
Fig. \ref{cr_Re_a2_n1_ALL_gam2_NEW_fig} about the vertical axis $\gamma_1=0$. Note that, for sufficiently large $\vert\gamma_1\vert$,
the critical curves for all values of $\gamma_2$ approach the axis $R=0$. So, this effect appears to be independent of $\gamma_2$.
Note also that the dependence of the critical curves on $\gamma_2$ (i.e. on what is happening at the outlet)
is relatively weak in comparison with the case of the pressure-stress
conditions (Fig. \ref{R_cr_a2_n1_ALL_s_fig}).

\begin{figure}
\begin{center}
\includegraphics*[height=8cm]{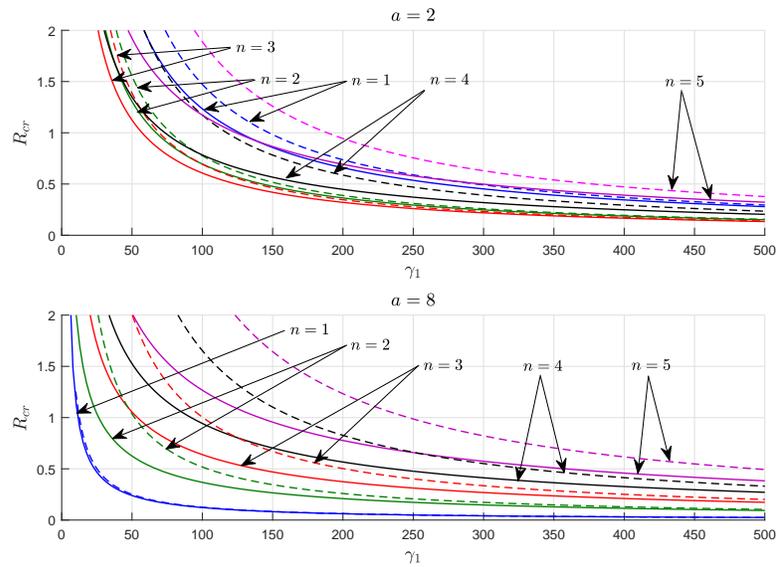}  %Fig 13
\end{center}
%\vskip -10mm
\caption{Diverging flow  with the pressure-no-slip conditions: critical $R$ versus $\gamma_1$ for large $\gamma_1$ for $a=2$ (upper plot) and $a=8$ (lower plot).
Dashed curves show the leading-order asymptotic values of $R_{cr}$ for $\gamma_1\gg 1$.}
\label{asymptotic_for_large_gam1_FIG}
\end{figure}

It turns out that it is possible to construct an asymptotic approximation of the eigenvalue problem for large $\gamma_1$. This is done in Appendix B, where
it is shown  that, at leading order, the critical values of $R$ are given by
\begin{equation}
R_{cr}=\frac{\Rey_{cr}(a,n)}{\gamma_1}+O\left(\gamma_1^{-1}\right) \label{4.34}
\end{equation}
where $\Rey_{cr}(a,n)$ is a certain critical value of the azimuthal Reynolds number, defined as
\begin{equation}
\Rey=\gamma_1 R = \frac{V_1^* r_1}{\nu}. \label{4.35}
\end{equation}
Here $V_1^*$ is the azimuthal velocity in the basic flow at the inner cylinder. The leading-order approximations computed using Eq. (\ref{4.34}), as well as
the critical curves obtained by solving the original eigenvalue problem, are shown as dashed and solid curves, respectively,
in Fig. \ref{asymptotic_for_large_gam1_FIG}. One can see that the dashed curves approach the solid curves as $\gamma_1$ increases, which indicates that the asymptotic formula (\ref{4.34}), obtained in Appendix B, works.
Note also that $Re_{cr}$ depends only on $a$ and $n$ and does not depend on $\gamma_2$,
i.e. on the azimuthal velocity at the outlet (of course, higher-order
approximations will depend on $\gamma_2$).
This means that, at leading order, the asymptotic behaviour of the critical curves
for $\gamma_2\neq 0$ is the same as that for $\gamma_2=0$. This explains our earlier observation that
the critical curves for all values of $\gamma_2$ approach the axis $R=0$ as $\gamma_1\to\infty$.

The fact that the flow becomes unstable at an arbitrarily small radial Reynolds number provided that $\vert\gamma_1\vert$ is sufficiently large
is quite unexpected because (in contrast with the case of the pressure-stress conditions) the basic flow is exactly the same as the one studied in \cite{IM2013b}.
The only difference
is that here the normal velocity condition at the outlet is replaced by the normal stress condition.
So, we can conclude that the normal stress condition drastically destabilizes the same viscous flow.

The asymptotic analysis of Appendix B has an interesting byproduct, namely: it turns out that the Couette-Taylor flow between
rotating cylinders (with a rotating impermeable inner cylinder and a non-rotating permeable outer cylinder) is linearly unstable to two dimensional
perturbations provided that the normal stress condition is imposed at the outer cylinder. This is in contrast with the
well-known fact that the classical Couette-Taylor flow (with the normal velocity condition at the outer cylinder) is stable to two-dimensional
perturbations. The instability
of the flow studied in the present paper at very low radial Reynolds numbers is a direct
consequence to this fact.

\begin{figure}
\begin{center}
\includegraphics*[height=8cm]{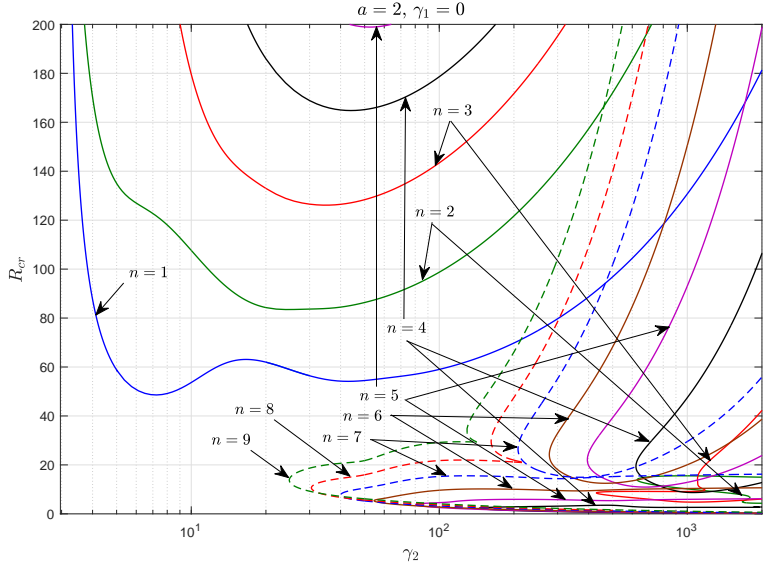}  %Fig 14
\end{center}
%\vskip -10mm
\caption{Converging flow with the pressure-no-slip conditions: critical $R$ versus $\gamma_2$ for $a=2$ and
$\gamma_1=0$.}
\label{conv_a2_asymp_suction_FIG2}
\end{figure}

\vskip 1mm
\noindent
\emph{Converging flow.}
Critical curves for the converging flows with the pressure-no-slip conditions for $\gamma_1=0$ and $a=2$ are shown
in Fig. \ref{conv_a2_asymp_suction_FIG2}. Curves in the top half of the picture for $n=1, \dots, 5$ are very similar to the critical
curves for the problem with the reference conditions. These curves are associated with the inviscid instability of section 2
and approach the vertical asymptotes $\gamma_2=\pm\gamma_2^{*}(a,n)$, where $\gamma_2^{*}(a,n)$ is the iniviscid instability boundary on the $(a,\gamma_2)$ plane, shown in Fig. \ref{stab_bound2}. For each azimuthal mode with $n=4, \dots, 7$, there are another two disjoint regions
where the corresponding modes are unstable. This is also true for the mode with $n=3$ but only one of the two regions is visible, as
the other is outside the range of $\gamma_2$ in Fig. \ref{stab_bound2}. For higher $n$, these two regions merge in to a single one (e.g.
the curves for $n=8$ and $9$).

\begin{figure}
\begin{center}
\includegraphics*[height=8cm]{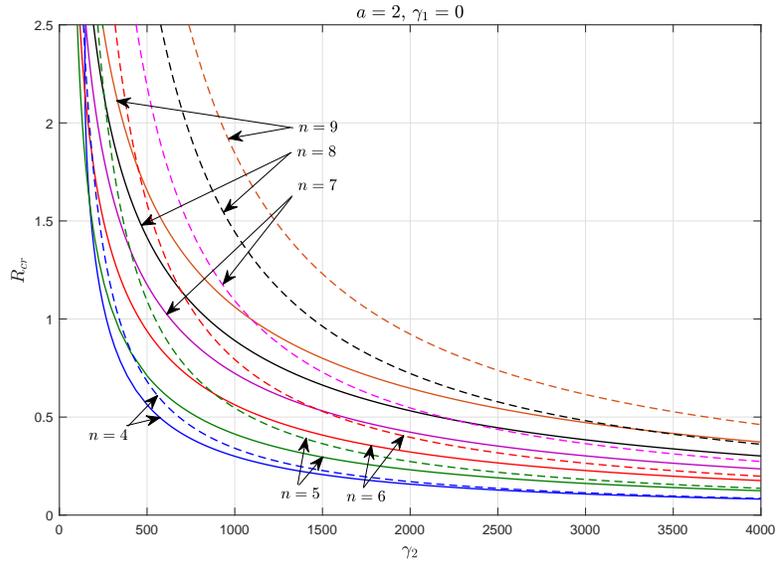}  %Fig 15
\end{center}
%\vskip -10mm
\caption{Converging flow with the pressure-no-slip conditions: critical $R$ versus $\gamma_2$ for $\gamma_2\gg 1$ for $a=2$, $\gamma_1=0$ and $n=4,\dots,9$ (solid curves). Dashed curves show the leading-order asymptotic values of $R_{cr}$ for $\gamma_2\gg 1$.}
\label{conv_a2_asymp_LARGE_GAMMA2_FIG}
\end{figure}

The instability regions that are in the right half of the figure,
but not too close to the horizontal axis $R=0$ are qualitatively similar to those discussed in \cite{IM2013a} and
associated with the instability of the boundary layer at the outlet (which reduces to the instability of
the asymptotic suction profile for $R\gg 1$). The instability regions that lie near the horizontal axis represent a new instability.
Figure \ref{conv_a2_asymp_LARGE_GAMMA2_FIG} shows that the lower boundary of these approaches $R=0$ as $\gamma_2\to\infty$. Asymptotic
behaviour of these curves for $\gamma_2\gg 1$ can be analysed in exactly the same manner as it was done for the diverging flows.
It is shown in Appendix B that, at leading order, the critical values of $R$ are given by
\begin{equation}
R_{cr}=\frac{\tilde{\Rey}_{cr}(a,n)}{\gamma_2}+O\left(\gamma_2^{-1}\right) \label{4.36}
\end{equation}
where $\tilde{\Rey}_{cr}(a,n)$ is a critical value of the azimuthal Reynolds number, defined as
$\tilde{\Rey}=\gamma_2 R = V_2^* r_1/\nu$ where $V_2^*$ is the azimuthal velocity in the basic flow at the outer cylinder.
The leading-order approximations given by Eq. (\ref{4.36}) and
the critical curves obtained by solving the original eigenvalue problem are shown as dashed and solid curves, respectively,
in Fig. \ref{conv_a2_asymp_LARGE_GAMMA2_FIG}. Evidently, the dashed curves approach the solid curves as $\gamma_2$ increases,
confirming that the asymptotic formula (\ref{4.36}) is correct.
Again, a byproduct of the asymptotic analysis is that
the classical Couette-Taylor flow with a rotating impermeable outer cylinder and a non-rotating porous inner cylinder
is unstable to two-dimensional perturbations if the normal stress condition is imposed at the inner cylinder.
This is even more surprising than the analogous result for the diverging flow because
it is well known that the classical Couette-Taylor flow is stable
even to three-dimensional perturbations when the inner cylinder is non-rotating \cite[see, e.g.,][]{Andereck,Iooss}.

\begin{figure}
\begin{center}
\includegraphics*[height=8cm]{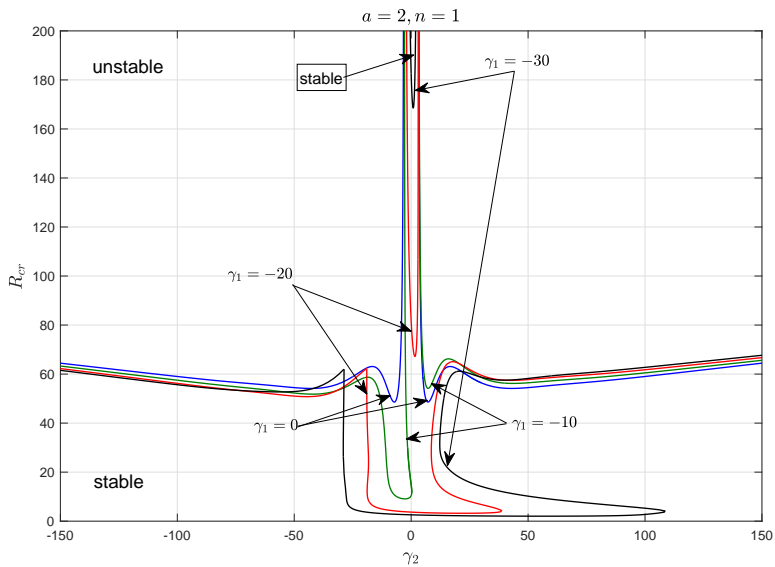}  %Fig 16
\end{center}
\caption{Converging flow  with the tangent velocity conditions: critical $R$ versus $\gamma_2$ for $a=2$, $n=1$ and $\gamma_1=0, -10, -20, -30$.}
\label{cr_Re_conv_tan_vel_ALL_gam1_FIG}
\end{figure}

Figure \ref{cr_Re_conv_tan_vel_ALL_gam1_FIG} shows critical curves for $a=2$, $n=1$ and $\gamma_1=0$,
$-10$, $-20$, $-30$.
As before,
the critical curves for the same $a$ and $n$ and for $\gamma_1=0$,
$10$, $20$, $30$ can be obtained by reflecting the curves in
Fig. \ref{cr_Re_conv_tan_vel_ALL_gam1_FIG} about the vertical axis $\gamma_2=0$.
The same feature as in
Fig. \ref{cr_R_a_2_n_1_ALL_s0_conv_fig} for the pressure-stress conditions appears here:
the curves for $\gamma_1=-10$, $-20$ and $-30$ cross the vertical axis $\gamma_2=0$.
For the converging flows with sufficiently large $\vert\gamma_1\vert$, there is a finite interval in $R$ where the flow
is unstable even for $\gamma_2=0$ (i.e. for purely radial flow at the inlet). For $\gamma_1=\pm 10$, the flow is unstable if
$R\in (10.38, 14.904)$; for $\gamma_1=\pm 20$, if $R\in (3.693, 81.635)$; for $\gamma_1=\pm 30$, if $R\in (2.588,182.578)$.
Again, for very high $R$ ($R\to\infty$), all critical curves approach the same vertical asymptotes as before
for all values of $\gamma_1$.

%%%%%%%%%%%%%%%%%%%%%%%%%%%%%%%%%%%%%%%%%%%%%%%%%%%%%%%%%%%%%%%%%%%%%%%%%%%%%%%%%%%%%%%%%%%%%%%%%%%%%%%%%%%%%%%%%%%
\setcounter{equation}{0}
\renewcommand{\theequation}{4.\arabic{equation}}

\section{Discussion}

We have shown that the instability of a simple steady inviscid flow found in \cite{IM2013a} also occurs if
the normal velocity condition at the outlet is replaced by the pressure condition. Moreover, under the pressure condition,
the flow is more unstable. We have also considered the stability of two families of steady viscous flows both of which reduce
to the single inviscid flow with the pressure condition in the limit of high radial Reynolds numbers. For both families,
all components of the velocity at the inlet, as well as the normal stress at the outlet were given. The only difference between them
was in the second condition at the outlet where either the tangent stress or tangent velocity were prescribed. As one would expect, both
families are instable due to the inviscid instability mechanism for sufficiently high $R$.
However, it turned out that, for moderate and small radial Reynolds numbers, the stability properties
(for both types of boundary conditions) may be very different from the results of \cite{IM2013b} obtained for the reference conditions.
In most cases, the pressure-stress and pressure-no-slip conditions have a strong destabilising effect, because
these conditions are less restrictive than the reference conditions (they allow perturbations with nonzero radial velocity
at the outlet).
In particular, in the problem with the pressure-no-slip conditions, both the diverging and converging flows turned out
to be unstable for arbitrarily small $R$, provided that the azimuthal velocity at the inlet
is much higher than the radial velocity. In these cases, we have derived asymptotic formulae for
critical radial Reynolds number which agree with numerical calculations.

\begin{figure}
\begin{center}
\includegraphics*[height=6.5cm]{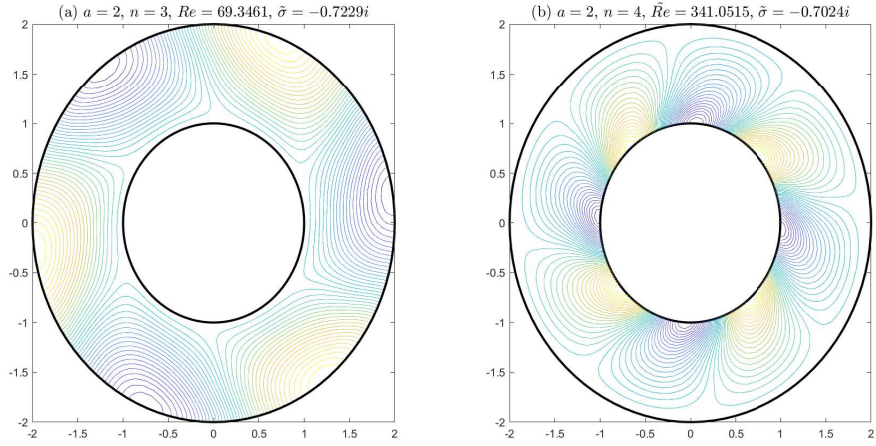}  %Fig 17
\end{center}
\caption{Neutral modes for the Couette-Taylor flow between a rotating impermeable cylinder and a stationary permeable
cylinder: (a) the outer cylinder is permeable; (b) the inner cylinder is permeable.}
\label{CT_neutral_eigs_FIG}
\end{figure}

As a byproduct of these asymptotic formulae, we
have found that two particular cases of the classical
Couette-Taylor flow, where one, \emph{impermeable} cylinder is rotating and the other, \emph{permeable cylinder} is stationary,
are unstable to two-dimensional perturbations provided that the normal stress condition (instead of the normal velocity condition)
at the permeable cylinder is imposed. This is in contrast with the well-known fact that
the Couette-Taylor flow is stable to two-dimensional perturbations. Moreover, in the case where the inner cylinder is non-rotating,
the classical Couette-Taylor flow is stable even to three-dimensional perturbations \cite[see, e.g.,][]{Andereck,Iooss}. The reason for
this instability is that the normal stress condition at a porous wall allows
nonzero flow into and out of the porous wall. In Fig. \ref{CT_neutral_eigs_FIG}, typical contour plots of the stream function
of neutral perturbations for both cases
of the Couette-Taylor flow are shown. Evidently, for these neutral modes the normal velocity at the permeable cylinder is nonzero.
Such modes are absent in the case of the reference boundary conditions.

Another interesting result, which is valid for both types of boundary conditions is that
there are flow regimes where the converging flows are unstable even if the azimuthal velocity at the inlet is zero
as on can see in Figs. \ref{cr_R_a_2_n_1_ALL_s0_conv_fig} and \ref{cr_Re_conv_tan_vel_ALL_gam1_FIG}.

The main conclusion of this paper is that boundary conditions at the outlet which include the normal stress condition
may completely change the stability of the flow. This is particularly apparent in the case of the pressure-no-slip conditions
where the basic steady flow is exactly the same as the one considered in \cite{IM2013b}, yet the change of one boundary condition at the outlet
makes the flow much more unstable for small and moderate values of $R$.

It was shown in \cite{IM2013b} that, in addition to an inviscid instability, there is another instability
in the problem with the reference boundary conditions and that it is related to the instability
of the boundary layer at the outlet. For problems considered here, we have not found
such instability in the case of the pressure-stress conditions. However, for the converging flows
with the pressure-no-slip conditions, for some azimuthal modes (with $n=3,\dots,7$), there are three different instability
domains (see Fig \ref{conv_a2_asymp_suction_FIG2}) where instability has different mechanisms: inviscid instability, instability of the
boundary layer at the outlet and instability due to the instability of the Couette-Taylor flow discussed above.
For the diverging flows with the pressure-no-slip conditions, Figs. \ref{cr_Re_tang_vel_cond_a2_FIG} and \ref{cr_Re_tang_vel_cond_a8_FIG}
show that
the instability domain already covers almost the entire $(\gamma_1,R)$ plane. Even if there were different mechanisms
of instability in different regions of the plane, it would be impossible to identify those regions.

There are many open questions in this area. Here we mention only one, perhaps, the most important question.
As was argued in section 3, the pressure-no-slip conditions may be relevant to flow between porous cylinders, provided
that the pressure in the porous cylinders is known. A more thorough approach would be
to consider flows in the free flow domain and in the porous cylinders and match them at the porous walls. A model of this type
has been considered by \cite{Tilton}, who have computed a steady viscous flow between a porous cylindrical membrane and
an impermeable cylinder with realistic boundary conditions, obtained using Darcy's law for the flow in the membrane.
As far as we are aware, there are no results on the stability
of steady flows between rotating porous cylinders with realistic boundary conditions, and this is a subject of a continuing investigation.

\vskip 2mm
\noindent
\textbf{Acknowledgements.} {The authors want to thanks Prof. V. A. Vladimirov for helpful discussions. A. Morgulis would like to acknowledge continuing support of the Southern Federal University (Rostov-on-Don).}

%\backsection[Declaration of interests]{The authors report no conflict of interest.}

%%%%%%%%%%%%%%%%%%%%%%%%%%%%%%%%%%%%%%%%%%%%%%%%%%%%%%%%%%%%%%%%%%%%%%%%%%%%%%%%%%%%%%%%%%%%%%%%%%%%%%%%%%%%%%%%%%%

\setcounter{equation}{0}
\renewcommand{\theequation}{A\arabic{equation}}

\section{Appendix A}\label{appA}

Here we show that the basic steady flows (\ref{4.12}), (\ref{4.13}) and (\ref{4.12}), (\ref{4.14}) are asymptotically
stable to
two-dimensional axisymmetric perturbations of arbitrary amplitude.
In particular, this means that if $n=0$,
then $\Real(\sigma) < 0$, i.e. the mode with $n=0$ cannot be unstable.

Let
\[
u=\frac{\beta}{r} + \tilde{u}(r,t), \quad v=V_{\beta}(r)+\tilde{v}(r,t)
\]
where $V_{\beta}(r)$ is given by (\ref{4.12}), and $\tilde{u}(r,t)$ and $\tilde{v}(r,t)$ represent an axisymmetric perturbation of finite amplitude.
Substituting these into Eqs. (\ref{4.17}), we obtain
\begin{eqnarray}
&&\left(\pr_{t}+ \frac{\beta}{r} \, \pr_{r}\right)\tilde{u} - \frac{\beta}{r^2}\, \tilde{u} -
\frac{2V_{\beta}}{r} \, \tilde{v} + \tilde{u} \tilde{u}_{r}  -\frac{\tilde{v}^2}{r}= -\tilde{p}_{r} +
\frac{1}{R} \left(L_0 \tilde{u}-\frac{\tilde{u}}{r^2}\right) ,  \nonumber \\
&&\left(\pr_{t}+ \frac{\beta}{r} \, \pr_{r}\right)\tilde{v} + \frac{\beta}{r^2}\, \tilde{v}
+\Omega_{\beta}(r)\tilde{u}  + \tilde{u} \tilde{v}_{r}  +\frac{\tilde{u}\tilde{v}}{r}=
\frac{1}{R} \left(L_0 \tilde{v}-\frac{\tilde{v}}{r^2}\right) ,  \nonumber \\
&&\left(r \tilde{u}\right)_{r} =0.  \nonumber
\end{eqnarray}
Here $\tilde{p}$ is the perturbation pressure and $L_0=\pr_{r}^2 + r^{-1}\pr_r$. The boundary conditions for
$\tilde{u}$ at the inlet are
$\tilde{u}\!\bigm\vert_{r=1}=0$ for the diverging flow ($\beta=1$) and
$\tilde{u}\!\bigm\vert_{r=a}=0$ for the converging flow ($\beta=-1$).
The incompressibility condition, together with these boundary conditions for $\tilde{u}$ imply that $\tilde{u}\equiv 0$, so that the first two of
the above equations simplify to
\begin{eqnarray}
&& - \frac{2V_{\beta}}{r} \, \tilde{v}   -\frac{\tilde{v}^2}{r}= -\tilde{p}_{r},  \nonumber \\
&&\left(\pr_{t}+ \frac{\beta}{r} \, \pr_{r}\right)\tilde{v} + \frac{\beta}{r^2}\, \tilde{v} = \frac{1}{R} \left(L_0 \tilde{v}-\frac{\tilde{v}}{r^2}\right) . \label{A0}
\end{eqnarray}
The second of these is independent from the first one and should be solved subject to appropriate boundary conditions for $\tilde{v}$,
while the first equation can be used to find the pressure $\tilde{p}$.

In the case of the pressure-no-slip conditions, we have
\[
\tilde{v}\bigm\vert_{r=1}=0, \quad \tilde{v}\bigm\vert_{r=a}=0.
\]
It had been shown in \cite{IM2013b} that Eq. (\ref{A0}) with these boundary conditions has only decaying (with time) solutions. This implies
the asymptotic stability.

Consider now the diverging flow ($\beta=1$) with the pressure-stress conditions, i.e.
\begin{equation}
\tilde{v}(1,t)=0, \quad \tilde{v}(a,t)- \frac{\tilde{v}(a,t)}{a}=0.  \label{A01}
\end{equation}
Let
\[
E=\int\limits_{1}^{a} \frac{\tilde{v}^2}{2} \, rdr.
\]
The equation of the balance of the perturbation energy, $E$, can be written as
\begin{equation}
\dot{E}+\left.\frac{\tilde{v}^2}{2}\right|_{r=a}+\int\limits_1^a\frac{\tilde{v}^2}{r} \, dr =-\frac{1}{R}
\int\limits_1^a \left(\tilde{v}_r-\frac{\tilde{v}}{r}\right)^2 r dr .
\label{A02}
\end{equation}
Equation (\ref{A02}) follows from the following chain of equalities
\[
\int\limits_1^a \left(\tilde{v}_r-\frac{\tilde{v}}{r}\right)^2 rdr=-\left.\tilde{v}^2\right\vert_{r=a}+\int\limits_1^a \left(\tilde{v}^2_r +\frac{\tilde{v}^2}{r^2}\right) rdr=-\int\limits_1^a  \left(\tilde{v}_{rr}+\frac{\tilde{v}_r}{r}-\frac{\tilde{v}}{r^2}\right)\tilde{v} \, r dr.
\]
[Here we used integration by parts and boundary conditions (\ref{A01}).]

For the converging flow ($\beta=-1$), the energy balance does not work so well, and we employ the perturbation angular momentum,
$\Gamma=r\tilde{v}$. In terms of $\Gamma$, Eq. (\ref{A0}) (with $\beta=-1$) takes the form
\begin{equation}
\left(\pr_{t} - \frac{1}{r} \, \pr_{r}\right)\Gamma  = \frac{1}{R} \, r\left(\frac{1}{r} \, \Gamma_r\right)_r. \label{A1}
\end{equation}
The boundary conditions for $\Gamma(r,t)$ that follow from (\ref{4.5}) and (\ref{4.9}) can be written as
\begin{equation}
\Gamma(a,t)  = 0, \quad \Gamma_r(1,t)- 2\Gamma(1,t)=0 .  \label{A2}
\end{equation}
Let
\[
M=\int\limits_{1}^{a} \frac{\tilde{\Gamma}^2}{2} \, rdr.
\]
After multiplying Eq. (\ref{A1}) by $r\, \Gamma$, integrating it from $1$ to $a$ in $r$ and performing
standard calculations involving integration by parts,
we find that
\begin{equation}
\dot{M}= - \frac{1}{R} \, \int\limits_{1}^{a}\Gamma_r^2  \, r dr -
\left(1+\frac{2}{R}\right) \, \frac{\Gamma^2(1,t)}{2}.  \label{A3}
\end{equation}
It can be shown that Eqs. (\ref{A02}) and (\ref{A3}) imply the inequalities
\[
\frac{\dot{E}}{E}\leq - C^+(a)\quad \hbox{and}\quad  \frac{\dot{M}}{M}\leq -\frac{1}{R} \, C^-(a),
\]
where $C^\pm$ are positive constants that depend on $a$ only. These estimates yield at least exponential decay of all perturbations.

%%%%%%%%%%%%%%%%%%%%%%%%%%%%%%%%%%%%%%%%%%%%%%%%%%%%%%%%%%%%%%%%%%%%%%%%%%%%%%%%%%%%%%%%%%%%%%%%%%%%%%%%%%%%%%%%%%%

\setcounter{equation}{0}
\renewcommand{\theequation}{B\arabic{equation}}

\section{Appendix B}\label{appB}

\emph{Diverging flow.} Consider the eigenvalue problem, given by Eqs. (\ref{4.24}), (\ref{4.25}), (\ref{4.31}) and (\ref{4.32}), in the limit $\gamma_1\to\infty$.
It follows from Eqs. (\ref{4.12}) and (\ref{4.16}) with $\beta=1$ that
\begin{equation}
V(r)=\gamma_1 V_0(r), \quad V_0(r)= - \frac{r^{R+1}}{a^{R+2}-1} +
\frac{a^{R+1}}{a^{R+2}-1} \, \frac{1}{r} + O\left(\gamma_1^{-1}\right). \label{B1}
\end{equation}
Let
\begin{equation}
\sigma=\gamma_1 \tilde{\sigma}. \label{B2}
\end{equation}
Substitution of (\ref{B1}) and (\ref{B2}) into Eq. (\ref{4.24})  yields
\begin{equation}
\left(\tilde{\sigma} +  \frac{in V_0}{r} \right) L \hat{\psi}-
\frac{in}{r} \, \Omega_{0}'(r) \hat{\psi} = \frac{1}{\Rey} \, L^2 \hat{\psi} + O\left(\gamma_1^{-1}\right)   \label{B3}
\end{equation}
where $\Rey$ is the azimuthal Reynolds number defined by Eq. (\ref{4.35}).

Now we make our key assumption that is consistent with the behaviour of critical curves in
Figs. \ref{cr_Re_tang_vel_cond_a2_FIG}--\ref{cr_Re_a2_n1_ALL_gam2_NEW_fig}, namely: $\Rey=O(1)$ as
$\gamma_1\to\infty$. With this assumption, the above formula for $V_0(r)$ simplifies to
\begin{equation}
V_0(r)=  \frac{1}{a^{2}-1} \, \left[\frac{a^2}{r} -r\right] + O\left(\gamma_1^{-1}\right). \label{B4}
\end{equation}
Note that if we discard the $O\left(\gamma_1^{-1}\right)$ term in (\ref{B4}), then $V_0(r)$ is the same as
the azimuthal velocity profile of the classical Couette-Taylor flow between rotating cylinders with radii
$1$ and $a$ in a particular case where the outer cylinder does not rotate, and the inner cylinder rotates
with angular velocity equal to $1$.

On substituting (\ref{B4}) into Eq. (\ref{B3}), we find that, at leading order,
\begin{equation}
\left(\tilde{\sigma} + \frac{in}{a^{2}-1} \, \left[\frac{a^2}{r^2}-1\right]\right) L \hat{\psi} =
\frac{1}{\Rey} \, L^2 \hat{\psi}.   \label{B5}
\end{equation}
At leading order, boundary condition (\ref{4.25}), (\ref{4.31}) and (\ref{4.32}) take the form
\begin{eqnarray}
&&\hat{\psi}(1)=0, \quad \hat{\psi}'(1)=0,  \quad \hat{\psi}'(a)=0,  \label{B6} \\
&&\frac{1}{\Rey}\left(\hat{\psi}'''(a) + \frac{1}{a}\, \hat{\psi}''(a) +\frac{4n^2}{a^3} \, \hat{\psi}(a)\right)
- \frac{2in}{a(a^2 - 1)} \, \hat{\psi}(a) = 0.  \label{B7}
 \end{eqnarray}
Thus, we have obtained the eigenvalue problem for $\tilde{\sigma}$.
It was solved numerically using the same method as the original eigenvalue problem. We found that
each azimuthal mode becomes unstable for $\Rey$ greater than some critical value $\Rey_{cr}$.
The critical azimuthal Reynolds
numbers for $a=2, 8$ and $n=1,\dots,5$ are shown in table \ref{tab:1}. Using these and formula (\ref{4.34}),
we plotted the asymptotic $R_{cr}$ as a function of $\gamma_1$
in Fig. \ref{asymptotic_for_large_gam1_FIG} (dashed curves).
\begin{table}
  \begin{center}
    \begin{tabular}{l|l|l|l|l|l|l|l|l|l|l}
      \multicolumn{1}{c|}{ } &\multicolumn{5}{c|}{$a=2$} & \multicolumn{5}{c}{$a=8$} \\
      \hline
      $n$ &1 &2 &3 &4 &5 &1 &2 &3 &4 &5 \\
      \hline
      $\Rey_{cr}$ &147.29 &77.58 &69.35 &117.36 &188.80 &12.54 &51.79 &100.48 &165.39 &246.62 \\
      \hline
      \end{tabular}
   \caption{$\Rey_{cr}(a,n)$}
   \label{tab:1}
  \end{center}
\end{table}

Now let us discuss the connection of this result with the classical Couette-Taylor flow.
Evidently, if boundary condition (\ref{B7}) were replaced by the condition $\hat{\psi}(a) = 0$
(i.e. no normal flow at $r=a$), this eigenvalue problem would coincide with the one arising
in the particular case of the classical Couette-Taylor flow mentioned above. It is known that
the Couette-Taylor flow is linearly stable to two-dimensional perturbations, although there seems to be
no formal proof of this fact \cite[see, e.g.][]{Drazin}. However, with boundary condition (\ref{B7}), the same flow
can be unstable. Recalling that the physical meaning of condition (\ref{B7}) is that the normal stress applied to the fluid
at $r=a$ is zero, we conclude that the Couette-Taylor flow with the normal stress condition at the outer cylinder
(instead of the normal velocity condition) is unstable to two-dimensional perturbations provided
that
\[
\Rey > \min_{n\in\mathbb{N}}\Rey_{cr}(a,n).
\]

\vskip 1mm
\noindent
\emph{Converging flow.}
Consider the eigenvalue problem, given by Eqs. (\ref{4.24}), (\ref{4.26}), (\ref{4.31}) and (\ref{4.33}), in the limit $\gamma_2\to\infty$.
The same arguments as for the diverging flow yield, at leading order, the following
eigenvalue problem for $\tilde{\sigma}=\sigma/\gamma_2$:
\begin{eqnarray}
&&\left(\tilde{\sigma} + \frac{in}{a^{2}-1} \, \left[1-\frac{1}{r^2}\right]\right) L \hat{\psi} =
\frac{1}{\tilde{\Rey}} \, L^2 \hat{\psi},   \label{B15} \\
&&\hat{\psi}(a)=0, \quad \hat{\psi}'(a)=0,  \quad \hat{\psi}'(1)=0,  \label{B16} \\
&&\frac{1}{\tilde{\Rey}}\left(\hat{\psi}'''(1) + \hat{\psi}''(1) + 4n^2 \, \hat{\psi}(1)\right)
+ \frac{2in}{a^2 - 1} \, \hat{\psi}(1) = 0.  \label{B17}
 \end{eqnarray}
Solving this problem numerically, we find that
azimuthal modes with $n=4,\dots,9$ become unstable for $\tilde{\Rey}$ greater than some critical value $\tilde{\Rey}_{cr}$.
The critical azimuthal Reynolds
numbers for $a=2$ and $n=4,\dots,9$ are shown in table \ref{tab:2}.
\begin{table}[h!]\label{tab2}
  \begin{center}
    \begin{tabular}{l|l|l|l|l|l|l}
     \hline
      $n$ &4 &5 &6 &7 &8 &9 \\
      \hline
      $\tilde{\Rey}_{cr}$~ &341.05 &546.34 &793.46 &1093.4 &1444.3 &1845.7 \\
      \hline
      \end{tabular}
  \end{center}
    \caption{$\tilde{\Rey}_{cr}(a,n)$ for $a=2$}
    \label{tab:2}
\end{table}

%%%%%%%%%%%%%%%%%%%%%%%%%%%%%%%%%%%%%%%%%%%%%%%%%%%%%%%%%%%%%%%%%%%%%%%%%%%%%%%%%%%%%%%%%%%%%%%%%%%%%%%%%%%%%%%%%%%

\bibliographystyle{jfm}
%\bibliography{jfm2esam}

\end{document}